\documentclass[10pt]{article}
\usepackage{amsmath}
\usepackage{amssymb}

\usepackage{graphicx}

\usepackage{cite}

\usepackage{color} 

\usepackage[labelfont=bf,labelsep=period,justification=raggedright]{caption}

\makeatletter
\renewcommand{\@biblabel}[1]{\quad#1.}
\makeatother

\date{}

\usepackage{url}

\begin{document}

\begin{flushleft}
{\Large
\textbf{From Social Data Mining to Forecasting Socio-Economic Crises}
}\\
Dirk Helbing$^{1,2}$ and Stefano Balietti$^{1}$ 
\\[3mm]
\bf{1} ETH Zurich, CLU, Clausiusstr. 50, 8092 Zurich, Switzerland
\\
\bf{2} Santa Fe Institute, 1399 Hyde Park Road, Santa Fe, NM 87501, USA
\\
$\ast$ E-mail: 
{\normalfont dhelbing@ethz.ch and sbalietti@ethz.ch}
\end{flushleft}

\section*{Abstract}

The purpose of this White Paper of the EU Support Action
``Visioneer''(see \url{www.visioneer.ethz.ch}) is to address the following goals:
\begin{enumerate}
\item Develop strategies to quickly increase the objective knowledge about social and economic systems
\item Describe requirements for efficient large-scale scientific data mining of anonymized social and economic data
\item Formulate strategies how to collect stylized facts extracted from large data set
\item Sketch ways how to successfully build up centers for computational social science
\item Propose plans how to create centers for risk analysis and crisis forecasting
\item Elaborate ethical standards regarding the storage, processing, evaluation, and publication of social and economic data
\end{enumerate}

\section{Introduction}\label{intro}

Modeling social and economic systems has been notoriously difficult in
the past. As a consequence, many of the grand challenges mankind is
facing in the 21st century are either of socio-economic nature or
involve human factors in a substantial way. Examples include

\begin{itemize} 
\item how to avoid socio-economic crises, systemic instabilities, and
  other contagious cascade-spreading processes,
\item how to design cooperative, efficient, and sustainable
  socio-technical and economic systems,
\item how to cope with the increasing flow of information, and how to
  prevent dangers from malfunctions or misuse of information systems,
\item how to improve social, economic, and political participation,
\item how to avoid ``pathological'' collective behavior (panic,
  extremism, breakdown of trust, cooperation, solidarity etc.),
\item how to avoid conflicts or minimize their destructive effects,
\item how to cope with the increasing level of migration,
  heterogeneity, and complexity in our society,
\item how to use environmental and other resources in a sustainable
  way and distribute them in the best possible way?
\end{itemize}

Many of these challenges cannot be solved by technology alone, but
require us to understand the collective social dynamics as roots of
these problems and key to their solution.  Despite the substantial
contributions that complexity science can make (and has already made)
in this connection, many of its implications have remained largely
theoretical so far. This is basically due to the lack of data, the lack of
computational power, and the lack of computationally tested
institutional designs in the past.  

This White Paper will primarily address, how the data gap may be
closed, while the EU project Visioneer \cite{Visioneer_site} will also
produce two further White Papers addressing the other two issues. We
expect that ground-breaking ICT research, in intense collaboration
with multiple other scientific disciplines, will in future be able to
identify the success factors of societies and to better address the
grand challenges of humanity.

\section{Strategies to quickly increase the objective knowledge about
  social and economic systems}
\label{sec_1}\label{strategies}

In the past, collecting data of human activity has been largely
obstructed by financial, technological and ethical issues. While the
ethical issues must be taken even more seriously in the future (see
Sec. \ref{ethical}), investments into experimental research and data
mining must be increased to reach the standards in the natural and
engineering sciences and to collectively benefit from the increasing
opportunities of collecting data and learning from them.  While
traditional research in the social sciences typically require many
years to generate relatively small data sets, new techniques to gather
data have recently become available. This includes lab experiments
\cite{Kagel,Bardsley,Friedman,Friedman2,Guala}, Web experiments
\cite{WebExp}, or the study of massive multi-player on-line games
\cite{Szell,FutureofSocialExperimenting,Bainbridge,Johnson}. Moreover,
some new technologies have created an abundance of new data, allowing
analyses with a previously unseen spatial and/or temporal resolution,
and also an analysis of the effects of the social environment or the
social interaction network,
etc. \cite{Christakis_obesity,Christakis_smoke,Christakis_happines,Christakis_cooperative,RealityMining,RealityMining_2}. New
terms such ``\textit{Information Cornucopia}'', ``\textit{Data
  Deluge}''\cite{DataDeluge} or ``\textit{Information
  Bonanza}''\cite{GuardianInformationPower} have been coined to refer
to the enormous amount of information produced by all these sources
(1200 exabytes\footnote{1 exabyte is $2^{60}$ bytes.} this year, only
150 exabytes in 2005).

In September 2008, the ``Jerusalem
Declaration''\cite{Jerusalem} stated: ``We are entering the era of a
high rate of production of information of physical, biological,
environmental, social and economic systems. The recording, accessing,
data mining and dissemination of this information affect in a crucial
way the progress of knowledge of mankind in the next years. Scientists
should design, explore and validate protocols for the access and use
of this information able to maximize the access and freedom of
research and meanwhile protect and respect the private nature of part
of it.''  Moreover, it is rightly pointed out that ``several
scientific disciplines once characterized by a low rate of data
production have recently become disciplines with a huge rate of data
production. Today a huge amount of data easily accessible in
electronic form is produced by both research and, more generally,
human activity.''

A short overview of data available to institutions, businesses,
governments, or everyone is given in Appendix \ref{Repository_Overview}.

A massive mining of socio-economic data or ``reality mining'' can have
quite substantial advantages (see Sec. \ref{crisisforecasting}):

\begin{itemize}
\item It can reduce serious gaps in our knowledge and understanding of
  techno-social-economic-environmental systems.
\item Crises Observatories (analyzing and mapping financial and
  economic stability, conflicts, the spreading of diseases...) could
  predict crises or identify systemic weaknesses, and help to avoid or
  mitigate impacts of crises.
\item Real-time sensing and data collection (``reality mining'' of weather data, environmental data, cooperativeness, compliance,  trust, ...) could reduce mistakes and delays in decision-making,
  which often cause an inaccurate or unstable system management.
\end{itemize} 

In the past, data about the spreading of diseases and the gross
national product, for example, became available with significant
delays (often weeks, months or years). Recently, however, it has been
discovered that the gross domestic product can be estimated in
real-time by measuring the light intensity at night, which can be
determined with satellite sensors \cite{GrowthFromSpace}. Similarly,
Google flu trends has been able to estimate epidemic spreading from
search requests \cite{FluTrends}. Furthermore, the ``wisdom of
crowds'' effect \cite{Surowiecki} is more and more used by crowd
sourcing approaches \cite{Howe,HoweBook,Wikinomics} such as prediction
markets to estimate future economic developments, outcomes of
elections, fashions, and socio-economic trends
\cite{Infotopia,Predictocracy}. These areas are now becoming an own
business branch, complementing classical consultancy, offering services like: real-time measurement of actual user activity, identification of trendsetters, opinion leaders, and innovators in social networks, trend prediction, trend tracking, etc. \cite{ICKN,PostRank}. They fit well
with other Internet-based applications such as eBusiness
\cite{Ebusiness,Ebusiness2} or eGovernance
\cite{Egovernance,Egovernance2}, for example. 

So far, it seems that the underlying technological innovations are 
mainly driven by businesses in the US. This, however, is partly a matter of legal regulations,
which give American companies a large degree of freedom regarding what
can be done with data of customers.

With the right institutional settings and the knowledge accelerator
proposed in another Visioneer white
paper\cite{VisioneerInnovationAccelerator}, Europe should be able to
catch up with the breath-taking developments in this area, which will
largely determine scientific and economic leadership in the future.

The potential of the newly available data has been also articulated by
leading scientific journals. Nature, for example, had a special issue
on Big Data, while Science hat an issue on Complex Systems and
Networks. Moreover, the National Science Foundation (NSF) has launched
a large ``Digging into data'' initiative, providing not only large
budgets, but also data (e.g. of science funding). A strategic program
on massive social data-mining, which was called for in {\it Nature}
\cite{Lazer}, is currently being worked out \cite{DigginIntoData}.

Scientific progress in the field requires a number of institutional
setting, which have to be either created or improved in the near
future \cite{Jerusalem}: 

\begin{itemize}

\item Supranational organizations, states, funding agencies and
  research institutions should recognize that information
  infrastructures are essential resources for the progress of the scientific knowledge, and
  are even more so in the future.

\item Supranational organizations should coordinate national projects
  collecting large databases to enlarge and harmonize the national
  databases in international information infrastructures.

\item Research societies should set up committees selecting protocols
  for the optimal access, use and dissemination of data produced by
  research activities or by the society.

\item States, funding agencies and research societies should take into
  consideration (i) the ``OECD Principles and Guidelines for Access to
  Research Data from Public Funding'' endorsed by the OECD Council on 14
  December 2006, (ii) the results of the 1997 CODATA studies on
  effective access to data for scientific research purposes in the
  natural sciences and (iii) the principles pertaining to research and
  statistical uses of data held by government agencies proposed at the
  Bellagio Conference in 1977.

\item For the promotion of access and use of data in scientific
  research, data produced while performing publicly funded scientific
  research should be made available to all scientists willing to use
  them, along with protocols describing the details of the data recording and data evaluation. 
  Scientists using data produced by other research groups
  need to acknowledge the origin of data. Failures to do so should
  be considered as scientific misconduct.

\item Scientists should organize the data produced by their research
  activity in a way, which makes the access to their data simple for
  other scientists. Funding agencies should support research
  consortia, Internet-based repositories and programs promoting data
  sharing both, by promoting these aspects in research grants and by
  special grants only devoted to these aspects.

\item Private companies producing large quantities of data should be
  given public incentives to make their data accessible to the
  scientific community. In order to protect intellectual rights and
  legitimate priority exploitation, data should be disseminated after a
  limited period of exclusive use (e.g. between 2 and 5 years).

\item Private companies operating under state license and producing
  large quantity of data should make their data available for research
  (in part or fully). When making a contract with a public institution, companies should be made contractually
  responsible for data dissemination (e.g. of traffic flow data or other recorded data of interest). They should make the data available after a reasonable
  period of time.

\item Also public entities (e.g. administrations) collecting and producing large quantities of data
  should make their data available for research, namely in a coded or surrogated form and, if restrictions of 
  transparency are in the public interest (e.g. for security reasons), after a sufficient period of time.
 
\item Research on data containing private or confidential information on individuals,
  groups or business organizations should be performed in a way that respects
  the privacy and protects the confidentiality of the agents investigated.

\item Data containing private information should be carefully managed
  when performing research projects, in order to prevent any illegal use or
  dissemination of the information or any abuse by third
  parties. Scientists need to take appropriate measures to ensure the
  best practice of managing such kind of information.

\item Public data with sensitive information should {\it not} be made
  publicly available for security reasons. However,
  it may still be in the public interest to provide surrogate versions of the datasets to researchers, which would 
  have to be prepared by specially experienced and authorized research groups for public research.

\item Supranational organizations, states, funding agencies and
  research institutions should promote the discussion about protocols
  and policies for the dissemination and processing of data, in order to increase the sensitivity for privacy issues
  among scientists and programmers such that the possibility of disclosure or exploitation of private information is avoided.

\end{itemize}

In Europe, Great Britain is already undertaking first steps along the above lines. In fact, in the
``Big Society Declaration'' \cite{BigSocietyDeclaration}, the UK
government stated its intention to implement a people's ``right
  to data so that government-held datasets can be requested and used
  by the public, and then published on a regular basis.'' This
ambitious promise is supported by continuous upgrades and
improvements of the UK government data Web site \cite{DataGovUk}. With the mission to unlock innovation, the British public sector
is making more and more  datasets available for download. For example, the
Combined On-Line Information System (COINS) gives free access to 
all UK Government expenditures. This decidedly innovative initiative, which
even goes beyond its American counterpart \cite{DataGov}, is taking the lead in Europe.

\section{Requirements for efficient large-scale scientific
  data mining of anonymized social and economic data}\label{requirements}

The study of complex systems (such as techno-social, economic, and environmental
systems) must take into account their sometimes counter-intuitive behaviors,
which implies that linear, conventional, and straight-forward modeling
attempts will usually be inappropriate. More advanced models 
(e.g. non-linear models with many heterogeneous interacting elements) require
more data to calibrate them, or to discover them with a data-driven approach. 
It is obvious that this task requires new computational approaches and the 
availability of large dataset of techno-socio-economic-environmental systems. 
In the social and economic sciences, data mining approaches have been largely 
lacking the the past. This has to change in the future, if we want to find better answers,
how to address problems such as financial instabilities and social conficts 
(see Sec. \ref{crisisforecasting} for further examples).
\par
\begin{figure}[htbp]
  \centering
  \includegraphics[scale=0.6]{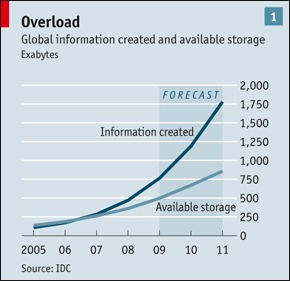}
  \caption{Projection of future information overload (from
    \protect{\cite{DataDeluge}}).}
  \label{overload}
\end{figure}

Fortunately, in the foreseeable future, scientists will have at
hand the volume and production rate of data needed to perform
meaningful analyses and even predictions. 
Back in 1890, at the Harvard Astronomical Observatory in Cambridge,
Massachusetts, large data-mining activities (for that time) were still
manually executed, who were scanning thousands of
photographic glass plates with the only aid of magnifying the glasses
\cite{HarvardComputers}.
Today, the scenario has dramatically changed. The pervasive
proliferation of small- and medium-sized computing devices into modern
society, and the steady influx of more people into the world's middle
classes during the last 20 years have enormously increased the total
number of persons leaving, producing and handling data every day. Nowadays, people have
great demand for knowledge and information, while information overload
is becoming a serious problem. This has pushed the overall
information production far beyond the available storage capacities (see
Fig. \ref{overload}). Ironically enough, even if they were able 
to store all the information currently being produced,
latency and bandwidth would become the new bottlenecks that would most
likely prevent us to \textit{read} what has been stored
\cite{RAID}. 

The current scenario is characterized by the following developments:
    
\begin{itemize}    
\item More and more decentralized data-sources are added everyday.

\item People continuously leave ``digital traces''
  \cite{Lazer} in their everyday lives.

\item Not only are there more data sources. The data generation rates are higher as well.

\item There exists a fast, pervasive network, allowing easy
  interconnection and data exchange among a growing number of data sources.
\end{itemize}
Given these trends, there is no doubt that petabyte datasets are
rapidly becoming a standard. However,
besides of the potential of these data for business and research, there are also
a number of related scientific challenges:
\begin{itemize}

\item Finding information in a targeted way becomes like seeking a needle in a hay stack.

\item The sheer number of such data sources, along with the amount of
  data produced, is shifting computation closer to the data sources,
  requiring streaming algorithms that produce decomposable and
  incrementally updatable results. 
 
\item An increasing number of applications is calling for a close
  integration of on-line and offline processing of data (e.g., process
  monitoring and control, algorithmic trading, real-time data
  warehousing, and others).

\item The wide heterogeneity of data representations makes any form of integrated processing extremely hard.

\item It is less and less clear what is valuable information and what is junk. The relevance of the content could even
change over time, as socio-economic or environmental conditions change.

\item More and more often, dataset or data collections contain personal and sensitive
  information.
\end{itemize}
These challenges require scientific breakthroughs, as
will be addressed in more detail throughout this White Paper. However,
it becomes visible already that, with an approach employing massive
data-mining, science can be pushed towards a new
methodological paradigm, which transcends ``the boundaries between
theory and experiments'' \cite{Muggleton}. In fact, while expert
knowledge will still be needed to judge how meaningful a certain model
is, starting from empirical data it may become possible to construct
scientific models in a semi-automated way (see \cite{VisioneerSocialComputing}). 
This development is already
progressing in meteorology, epidemiology, and systems biology, and
soon a data-driven approach will be also possible in the 
socio-economic sciences. This would require an ICT system capable of
combining various facts, logical reasoning, and of assessing the
relevance of outcomes. Such a system would not only have to include
the state-of-the-art modelling techniques. It would also have to be
able to systematically study implications of different possible
interpretations of the data.
Furthermore, it should also be able to determine the reliability of the
respective conclusions, considering the statistical validity of the
underlying models in order to estimate the likelihood that certain
scenarios would happen.
\par
So far, an integrated approach to techno-socio-economic-environmental
data mining seems to be missing. In order to change this, it will be
necessary to 
\begin{enumerate}
\item efficiently transfer existing and future data-mining knowledge from other research areas into the social and economic sciences,
\item build up a distributed data center and a multi-disciplinary international community of
scientists participating in its activities (as outlined in Sec. \ref{centers}),
\item develop tools to extract significant patterns and relationships from large datasets (see Sec. \ref{stylizedfacts}),
\item elaborate suitable visualization tools for large dataset, be
  they of empirical nature or results of large-scale computer
  simulations (see \cite{VisioneerSocialComputing}),
\item address the ethical, legal, and institutional issues related to these activities (see Sec. \ref{ethical}). 
\end{enumerate}

\section{How to collect stylized facts extracted from large data sets}\label{stylizedfacts}

In order to stimulate the modeling and computer simulation of social and economic systems,
it would be important to have a collection of stylized facts of these systems, i.e. a book of all the empirical observations that have been consistently made in a number of contexts. The current lack of
such collections seriously impedes the progress in the area of developing realistic computer models. In 
order to create such lists, the following measures can be taken:
\begin{itemize}
\item One could employ crowd sourcing methods by creating specialized Wikis or by organizing so-called
``Hilbert workshops''. Such workshops would be multi-disciplinary forums to assess the existing empirical evidence
for or against theoretical pillars in economics and the social sciences (pretty much the same as the famous
Copenhagen workshop addressed the controversies of quantum mechanics). Hilbert workshops would also
determine knowledge gaps, future scientific challenges and promising research routes to address them. 
Results of such crowd sourcing approaches should be summarized from time to time by review papers or books.
\item Targeted laboratory or Web experiments could be performed to study behavioral regularities under well-controlled
conditions.
\item Pattern recognition and machine learning approaches could be applied to identify significant
relationships in an automated way.
\end{itemize}
Some people think that the lack of collections of stylized facts in the socio-economic sciences may be a result
of the complexity of these systems and the non-existence of such facts due to the great degree of flexibility of
social and economic systems. However, there {\it are} actually a number of stylized facts, as the following list indicates: 
\begin{enumerate}
\item Moore's law (according to which the number of transistors that can be placed inexpensively on an integrated circuit is doubling approximately every two years) \cite{Moore},
\item the Fisher equation of financial mathematics (which determines the relationship between nominal and real interest rates under inflation) \cite{Fisher},
\item the fat tail character of many financial and economic
  distributions \cite{Mandelbrot,Fama},
\item the Mattew effect (rich-gets-richer effect) \cite{MatthewEffect,BarabasiScaling}
\item Dunbar's number (limiting the number of people one can have stable social relations with) \cite{Dunbar},
\item Pareto's principle (according to which roughly 80 percent of an effect comes from about 20 percent of the causes) \cite{Pareto},
\item Zipf's law (determining the distribution of city rank sizes and many other things) \cite{Zipf},
\item the gravity law (describing the distribution of trade flows and migration) \cite{GravityLaw_Ravenstein,GravityLaw_Zipf,GravityLaw_Tinbergen},
\item Goodhart's law (according to which any observed statistical regularity, e.g. a risk model, breaks down once pressure is placed upon it for control purposes ) \cite{Goodhart,Goodhart_New,Goodhart_Danielsson}.
\end{enumerate}
We do not want to imply here that the above ``laws'' would {\it always} be true, but still they are quite useful to understand certain behaviors of socio-economic systems. Moreover, it should be pointed out that there are many more recurrent observations that would qualify as stylized facts, i.e. abstracted, generalized, simplified, or idealized features that are characteristic for certain aspects of social or economic systems. 
\par
A confusing feature of social and economic systems is the fact that some characteristics are true under certain circumstances, but other, often incompatible or contradictory characteristics apply under other circumstances. This can be caused by parameter- and/or history-dependencies of the behavior of social and economic systems (or complex systems in general). Some of this confusion can be probably resolved by representing such dependencies through phase diagrams, considering the possibility of multi-stability \cite{PluralisticModeling}. This, however, requires an accurate description not only of the different types of behaviors observed in social and economic systems, but also of the conditions under which they occur. We anticipate that massive data mining can lead here to a substantial advances and breakthroughs. 
\par
In fact, the emerging ICT-based ability to collect stylized
facts from massive datasets has been celebrated as the
``fourth paradigm of science'', after theory, experiments and
simulations \cite{Fourth_Paradigm}. There is a growing literature
supporting this perception. For example, Banko and Brill \cite{Banko_Brill} have
shown that, given a sufficiently large amount of training data, the
accuracy of different machine learning algorithms eventually
converge. Another paper by three Google researchers
\cite{UnreasonableData} argues that the ``biggest success in
natural-language-related machine learning have been statistical speech
recognition and statistical machine translation'', just because
``large training set of the input-output behavior [\ldots] were
already available \textit{in the wild}''. Brant et al. \cite{Brants}
even demonstrated how, in natural language models, a worse technique
called ``stupid backoff'' could outperform the state-of-the-art
approach, if opportunely trained on a large enough datasets (two
  trillion words). This is, because we have such large
data sets today that traditional algorithms are no longer efficient enough
for mining and analysing \cite{ScalabilityDataMining}. Sophisticated
and computationally expensive models can simply not cope anymore with
the \textit{web-scale} amount of data.

Massive datasets are clearly beneficial for science and they will
facilitate new scientific breakthrough as they will lead to better
algorithms and systems capable of overcoming the problems of
traditional data-mining approaches. We will review a few here.

\subsection{The way Google does it: MapReduce and distributed storage systems}
\label{Hadoop}

Cory Doctorow in his Nature article ``Welcome to the Petacenter'' says
``you can't go far in this world without some sort of comparison with
Google'' \cite{WelcomePetacenter}. This is absolutely true when it
comes to data center operations, which Google keeps very
secret. Nonetheless, during conferences or other special events, it is
possible to get an insight of company's internal organization
\cite{GoogleSpotlight}.

The core of modern computational datamining infrastructures relies on two cardinal
pillars: (i) a software to store the data over thousands of machines;
(ii) a software to retrieve and perform computation with data spanned
over thousands of machines. The former is accomplished by two main
components: a special File System (GFS) and a special BigTable, to
guarantee fault-tolerant access respectively to unstructured and
structured data. The latter is achieved through a special program
called MapReduce.

MapReduce\cite{MapReduce} is a patented implementation of the
famous \textit{Divide and Conquer} algorithm design paradigm in
computer programming. It is aimed to support distributed computing on
large data sets on clusters of computers.

In a nutshell, the MapReduce approach takes a computational problem
and ``maps'' it to different nodes that will take care of solving a
part of it. Moreover, since they are {\it also} running an instance of
MapReduce, they can split their task into sub-tasks for distribution
in the computer cluster as well, and so on. Once the single computational tasks are
completed, they get ``reduced'' to a single output and returned back to
the initial caller. There exists also a specific language
on top of MapReduce called Sawzall \cite{Sawzall}, which
exploits the parallelism inherent in having data and computation
distributed across many machines.

Apparently, MapReduce is so fundamental that it was not released as open source, 
in contrast to what has been common for similar developments so far.  
Nonetheless, the concept behind MapReduce inspired
Hadoop, which is a free implementation released in 2009.
In certain distributed contexts, the fact that Hadoop runs in JAVA can
represent a 
serious disadvantage, since it may create a coding overhead as compared to other languages. Hence, PIG latin, a specific high-level language for expressing data analysis
programs in parallel environments, has been developed for the same reason
for which Sawzall was created.
 
\subsubsection{Hadoop}

Since Hadoop is a kind of open implementation of MapReduce, we will focus
on this software system, although most of the considerations apply to both of
them.

Hadoop is a scalable fault-tolerant grid operating system for data
storage and processing \cite{Awadallah}. The advantage of its
programming model is that it provides a powerful abstraction of
\textit{what} to do and \textit{how} in the large-scale processing of data. 
This is accomplished by applying the MapReduce paradigm on
unstructured and structured data, over a self-healing high-bandwidth
clustered file system (HDFS).

``The speed of Internet has not kept pace with the growth of
scientific datasets''. Therefore, today, and most likely in the future as well,
we cannot copy a petabyte database to our personal
workstation. Rather, the only viable solution is ``sending the
computations to the data''\cite{ScienceExponential}. Within Hadoop,
computing is actually moved to the data. Even more impressively, this
is achieved with linearly scaling performance.

The following list gives examples what Hadoop is suitable and currently being used
for \cite{Awadallah,Hofman}:
\begin{itemize}
\item parsing and manipulating large text collections,
\item building the Web search index,
\item processing news/content feeds,
\item content/ad targeting optimization,
\item classifications (e.g. fraud and spam and detection) and
  clustering,
\item lexica of social networking Web Sites: trends of words on walls,
\item collaborative filtering,
\item batch video/image transcoding,
\item gene sequence alignment,
\item individual node/edge-level (e.g. clustering coefficient,
  breadth-first search, etc.) calculations.
\end{itemize}
Nevertheless, Hadoop is still not the definitive answer to all kinds of storage-processing problems. 
At present, relational databases are still the optimal
solution when transactional consistency, schema-dependency and
a highly interactive response to complex queries are required up to hundreds of terabytes.

\subsection{Real-time knowledge mining}

Until now, organizations have followed the paradigm to gather data
first, and to analyze them later. However, this paradigm falls short in view of the
contemporary challenges, which require for real-time reactions to changes
in the respective environment. This calls for a new generation of data mining and knowledge generation.

This is a very active research direction in data mining and it may be
organized around three main themes:

\begin{enumerate}
\item \textit{Continuous data analysis over streaming data:}
  Techniques for the on-line monitoring and analysis of high-rate data
  streams \cite{DM_Stream}. Part of this theme is to develop
  mechanisms for handling uncertain and inexact
  data \cite{Aggarwal_2009_uncertain_data}.

\item \textit{On-line ingestion of semi-structured data and
  unstructured ones} (e.g., news items, product reviews, etc.), which
  is important to allow one to profit from the wealth of information
  outside the realm of structured data, both inside and outside an
  organization \cite{Tsytsarau_2011,Hoffart_2011}.

\item \textit{Real-time correlation of streaming data (fast streams)
  with slowly accessible historical data repositories:} Mechanisms for
  correlating and integrating real-time data (fast streams) with
  historical records (usually stored in large and slow-to-access
  static data repositories), in order to deliver a contextualized and
  personalized information space. This adds considerable value to the
  data, by providing (historical) context to new data
  \cite{Polyzotis_2008,Camerra_2010}.
\end{enumerate}

The three above-mentioned research themes are not to be seen separate from each other.
Exploring the interactions among them could leverage large positive
externalities due to their synergistic roles.

\subsection{Reality mining: capturing collective and social behaviours
in an ubiquitous ICT society}

Capturing human collective and social behaviours on a large scale
can play an important role in the response to unexpected
events such as crises or disasters, and for the support of
populations in difficult situations. Deviations of the dynamical system from
nominal states may become key predictors of situations requiring
societal and individual adaptation. While capturing the state of the complex
dynamical system and its changes on a large scale and with a sufficiently fine-grained
temporal and spatial resolution was impossible in the past, it now becomes
feasible to understand social and
collective dynamics in humans systems with a resolution unseen before. This will allow one
to design real-time large-scale societal surveys, to facilitate eGovernance solutions, and to create
support and response strategies for adverse events.

Reality mining projects would be crucial:

\begin{itemize}

\item to increase our understanding of collective and social behaviours
  through larger datasets and

\item to refine social and cognitive models of human behaviour,

\item to devise individual or societal response strategies in order to assist and help
coping with major challenges; examples would be
\begin{itemize}
\item emergency, evacuation and crisis management operations (real-time collective
behaviour sensing can be used to inform individuals about appropriate actions and suitable relief operations),

\item well-being and painless ageing (changes in the of social
  dynamics may indicate the onset of depression, which may be counter-acted by
  appropriate social support strategies),

\item shaping the metropolis of the future (knowledge of the collective
  behavior in existing urban spaces could provide data required for better future designs),

\item creating a ``multinational adapter'' or ``multi-cultural guide'' (a real-time ICT-enabled
  translator for verbal communication and non-verbal expectations like cultural values and norms,
  to lower communication and cultural barriers). 
  \end{itemize}
\end{itemize}

\subsection{Some problematic issues of data mining}

Having talked about the virtues of machine learning and automated pattern recognition before, we would like
to end this section with a word of warning. The following problems illustrate that one must be aware of the limitations of such approaches, and the use of complementary approaches, such as human intuition, experimental studies, and computer simulations \cite{VisioneerSocialComputing}, appears to be absolutely necessary to avoid potentially serious consequences (see also Refs. \cite{Econophysics,PluralisticModeling}):
\begin{itemize}
\item \textit{Na\"{\i}ve data mining and wrong models:}
Large datasets support the temptation to perform ``blind'' data mining, i.e. to apply pattern recognition
tools and statistical methods without a deeper understanding of the underlying assumptions and implications
of the results. Most algorithms deliver numerical results under a wide range of conditions, even when their applicability is not given or questionable. Typical problems of data mining are
\begin{enumerate}
\item (mis-)interpreting correlations as causal relationships,
\item collinearity of variables, leading to non-unique parameter fits and more or less arbitrary parameter choices,
\item underfitting or overfitting of mathematical models,
\item ignorance of underlying simplifications or critical approximations.
In fact, a large fraction of statistical conclusions are
scientifically invalid \cite{ScienceInvalid}.
\item Large amounts of data (BigData) can produce the illusion of control over
complex systems. This is quite dangerous and probably 
what happened during the recent financial crisis. Banks and rating
agencies were feeding their models with enormous amounts of data, but the upcoming crisis 
was not noticed, as the models did not reflect reality well enough. This was maybe ``the first crisis to be
sparked by big data'' \cite{EconomistDataDataEverywhere}.
\end{enumerate} 

\item \textit{Information pollution:} Usually, there is no guarantee that data in the Internet would be in any way reliable. In fact, many comments and blogs have a relatively low quality, and co-creation platforms such as Wikipedia have to cope with a considerable level of manipulation attempts and ``vandalism''. Rumors about other people and companies may be placed to promote self-interest (e.g. pursue marketing strategies), and the reliability of their contents is often questionable. Viral marketing strategies make it even more difficult to distinguish facts from fake. It is also known that some users try to manipulate page ranks and reputation values by using multiple identities, robots and other tricks. Further problems result, as the Web is dominated by a majority of ``time-rich'', rather than experts. For example, more and more users upload contents to the Web, most of which are not the results of specialized skills. Some contents may even be intentionally misleading. Therefore, the quality and representativeness of data on the Web is a problem, and ``information pollution'' is an issue that will have to be addressed by future data mining attempts. Another problem is the herding effect (and the related multiplication of certain kinds of information), which easily occurs in information-rich environments due to a lack of overview and a desire to find hints for orientation. 
\end{itemize}

\section{How to successfully build up centers for computational social science}\label{centers} 

Data-intensive processing is beyond the capability of any individual
machine and requires clusters of hundreds or even thousands of computers. At
the same time, data is also becoming increasingly more complex, and
call for new advanced visualization techniques to support the
intuition of scientists in the hypothesis generation process. In this field, the
new advances in virtual reality can play a vital role.  The
augmented complexity of the data requires the analysis to become more
and more a collaborative task. 

Applied computer science is gradually taking the role which
mathematics had until the end of the last century. That
consisted in: ``providing an orderly formal framework and explanatory
apparatus for the other sciences''\cite{ScienceTwoWays}. As a direct
consequence, more and more fundamental experiments require computers
and customized software as essential components of their design. They
crunch large datasets and eventually yield more data, the analysis of which
is complex and increasingly more difficult to
reproduce \cite{ScienceExponential,CommunityCleverness}.

In order to avoid the risk of confining computational social science
behind the curtains of private data with no replicability of
experiments and methods, one needs to set up a common environment
establishing legal rules, ethical and technological requirements 

\begin{itemize}

\item to define interoperability standards for data archiving,
\item to settle protocols for documenting instruments, procedures and
  measurements, and
\item to guarantee the security and privacy of the data

\end{itemize}
(see Sec. \ref{ethical}). 

An institutional approach to establish such a framework is encouraged
by many stakeholders \cite{MatterTrust,Butler,HowDataGrow}, and benefits
would be manifold and immediate. In particular, community standards
would facilitate the data reuse by ``making easier to import, export
combine and understand data'' in a safe environment. They would also
``eliminate the need for each data creator to develop unique
descriptive practices'' \cite{HowDataGrow}.

However, as Lazer et al.\cite{Lazer} pointed out, the privacy issue on
data archiving and disclosing must be handled with special care, since
``a single dramatic incident involving a breach of privacy could
produce rules and statutes that stifle the nascent field of
computational social science''. A well-known incident, which
revealed the identities and surfing behaviors of a large number of
users, is the publication of a paper by the research division of a large Internet 
provider \cite{AOLIncident}, to which a not completely anonymized
version of the dataset used for their study was attached. More recently, a one million dollar prize competition for developing a better recommender algorithm was cancelled, because two scientists from the University of Texas showed that it was possible to easily de-anonymize the training dataset \cite{Deanon}. 

In order to prevent new incidents from happening and to foster the use of a 
privacy-preserving data mining, we envisage an ethical framework, which would maintain the
research potential of a data-rich socio-economic science, while
protecting data privacy (see Sec. \ref{anonyrandom}). At the same time,
mechanisms (e.g. special reputation points) are required to reward
researchers \cite{ScienceExponential,CommunityCleverness} sharing
their data (without compromising their ownership). For example, new
scientific performance indices may allow scientists to profit from the
citations of all publications using their data, which would require to
create a unique way of citing datasets. The Dataverse Network Project
\cite{DataVerse} is a first attempt to enable data archiving based on
standards and exchange protocols, rewarding data owners through data
management and persistent citations. Other Web data-sharing
initiatives have already been launched, with different focuses and
targets. For example, Many Eyes wants ``to provide masses with access
to visualization tools, especially interactive ones''
\cite{NatureDataSharing}. Swivel instead allows easy dataset
composition and reuse. For references see Appendix
\ref{Repository_Overview}.

Finally, defining clear rules in a protected data-free environment as
such would foster innovation by stimulating different factors,
e.g. multidisciplinary collaborations among scientists, the 
re-use of the data, secondary and tertiary data processing, critical
re-analyses, etc. Results will thus cumulate more rapidly in multiple 
disciplines and increase the knowledge
needed to build up new models that are necessary for the understanding of
modern techno-socio-economic-environmental systems \cite{Ostrom}.

\subsection{Why and how to get computation to the data}

Lin and Dyer \cite{LinDyer} have recently published a list of
``\textit{Big Ideas}'' that are essential for the realization of big data centers. 
We will report and comment them here, integrating contributions of other authors as well.

\subsubsection{Scaling out}

While ``scaling up'' is a term used to describe a technical upgrade of the
  processing power of a data center using high-end servers, ``scaling out''
 is used to describe a strategy using a larger number of low-end servers.
 Obviously, the latter also means that one needs to design the
  applications such that they can efficiently run in a highly distributed
  environment. This is exactly what software based on the MapReduce
  approach, e.g. Hadoop, is for.

Scaling out has a definitive advantage: it is cheaper. In fact, the
cost of scaling up does not scale linearly, i.e. machines twice as
fast will most likely be notably more than double priced. On the other
hand, low-end servers are served by a very competitive market with
a large economy of scale. While the higher costs of scaling up would be
justified by significantly higher performances, these are actually not 
reached. In fact, in 2007 scaling up has been compared to
scaling out in the context of search applications for the Web \cite{ScaleOut}. The
conclusion was ``that scale-out solutions have an indisputable
performance and price/performance advantage over scale-up for search
workloads''. In 2009, Barroso and H\"olze \cite{ScaleOut2}, after
testing both approaches under several workload configurations have
declared that the huge costs of the scaling-up solution does not find
a justification in clearly superior benchmarks. This fact speaks clearly for a 
grid or cloud computing concept.

\subsubsection{Reliability on software} 
\label{hw_failure}
Hardware failure must be assumed as a fact of large 
computer clusters. Certifications and periodic controls of the hardware are
necessary, but not sufficient to avoid hardware faults: In a data
center with thousands of machines at least one failure per day is
inevitable. This is why big companies have started implementing
reliability at the level of the application for many years
\cite{GoogleSpotlight}.

\subsubsection{Move processing to the data} 
\label{local_computing}

Traditionally, storage and processing of data has been kept
separated for technical and logical reasons, but this approach is no
longer efficient for dealing with huge amounts of data. In fact, storage filers can become a significant bottleneck when large chunks of data need to be copied to the grid for processing. Problems appear to be even more acute when dealing with raw-data and their transformation into structured information \cite{Awadallah}, namely for the following reasons:

\begin{itemize}

\item \textit{Data errors and reprocessing}. Data errors may be
  encountered even a long time after the original processing. 
  Given this and the prohibitive costs to reprocess tape-data, raw
  data have to retain on-line for extended periods.

\item \textit{Conformation loss}. Conversion from a raw format to
  structured information implies a certain degree of information
  loss. Again, this requires to retain the raw data for longer time periods.

\item \textit{Ad-hoc queries on raw data}. Running ad-hoc queries on
  storage filers is not possible, since they only store and cannot
  compute.

\item \textit{Schema pre-processing}. It is usually not possible to
  process unstructured data, which need a pre-processing stage in
  order to fit into a predefined schema. With huge amount of data,
  such a task is increasingly causing large delays in processing.

\item \textit{Limitations of access}. Storage filers often do not offer an
  optimal solution to access and process data from any branch of the
  organization.

\end{itemize}

A hybrid ``Storage-Compute'' solution with distributed file system
(e.g. such as Hadoop) is able to overcome all the above issues by
coupling processors and storage.

\subsubsection{Organize and process data sequentially}

Notwithstanding predictions that hard-disks will replace tapes, 
and solid-state drives may replace hard-disks, 
rotating media storage devices are still the unchallenged standard for
data-warehousing solutions and this dominance is not going to be
undermined in the near future.\footnote{Solid-state \textit{flash} memories
have incredibly increased life-cycle and capacity \cite{SolidState}, but their cost per megabyte is still prohibitive compared to traditional drives.}
  
In rotating devices, the seek operation takes more time than any other
part of the input/output process. In these, accessing data sequentially is
orders of magnitudes faster (up to 150.000 times faster) than accessing them 
randomly. Moreover, even random access to memory may be slower than sequential access to disks \cite{PathologyBigData}. While an efficiency loss in accessing information may still be acceptable for normal data collections, ``Big Data greatly magnifies the performance
impact of suboptimal access patterns''\cite{PathologyBigData}. In spite of the steady increase in required 
storage space, it is therefore largely desirable to organize the data on disks,  
and to do all the data access in a sequential order at any level of the hierarchy of processing stages.

\subsubsection{Redundancy can improve efficiency and reliability}

Data replication is a good shield against single-node hardware
failures (see \ref{hw_failure}), but it has also another important
advantage. A major problem of parallel computing is how to uniformly
distribute the workload across the nodes. In fact, when the hardest
part of a job is executed on a single machine, that node will become
the bottleneck of the whole computational task, thereby losing all the
advantages of parallelism. This scenario is not infrequent, and its
occurrence depends (i) on the type of analysis performed with the data
and (ii) on how the data is ordered and aggregated locally (see Sec.
\ref{local_computing}). In fact, whether data are aggregated according
to the time stamp or another suitable identifier (index) can transform
a local task to a distributed one and viceversa. Given sufficient
storage space, it is possible to replicate the data in the cluster
such that the gain of the distributed approach is not lost, regardless
of the type of analysis performed.

\subsubsection{Service oriented architecture}

In the last decade, the software industry has moved giant steps in the
direction of \textit{integration}. Business-to-business transactions,
as well as Web mash-ups have flourished thanks to the great
proliferation of public api (application programming interface), and
Web services. Behind all this stands the idea of a Service
Oriented Architecture (SOA).

SOA is actually not a new concept. Service is in fact just another
name to indicate large-scale components wrapped behind a standard
interface. The term architecture instead is referring to previous
ideas such as software bus, IT backbone, or enterprise bus.

Probably, the real novelty of SOA is the rigorous use
of standard interfaces, mostly Web services, through which it is possible
to automatically develop local components capable to exchange data
with remote counterparts, regardless of the operating platform and
programming language.

It basically added a transparent layer on top of standard
components, which has pushed a large part of the complexity of
programming into its underlying infrastructure (containers, middleware
stacks, etc.), hiding the details of the implementation to developers,
who in this way can focus on the application.

Apache WSIF (Web Service Invocation Framework) \cite{WSIF} probably
represents the state-of-the-art of service-oriented architecture. In
fact, it allows a completely dynamic service invocation, allowing
plugging-in and choosing new bindings at run-time.

Such a solution can incredibly extend software life-time, as it
supports debugging, enhances scalability, and promotes
integration. For an average software company only 20\% of the costs
are strictly related to software production, while about 80\% result
from its maintenance, e.g. updating and integration. This clearly
shows how beneficial the approach of SOA and WSIF can
be. Furthermore, insisting on such concepts will eventually give birth
to next-generation data centers. They will be so seamlessly
integrated, that it will be possible to move jobs from one data center
to another, more or less automatically \cite{GoogleSpotlight}.

\section{How to create centers for risk analysis and crisis forecasting}\label{crisisforecasting}

Weather forecasts are perhaps the most acknowledged example of
successful prediction of the behavior of a complex system, even 
though restricted to short time scales. They are
the final product of the relentless integration of sensor data and
historical patterns into computational simulations, which are performed by 
supercomputer centers. Weather forecast can estimate the trajectory and
intensity of hurricanes or extreme weather and possibly save many lives 
as well as large economic values in agriculture. They
unmistakably show us that is not an utopia to predict, within a certain
degree of approximation, the behavior of complex systems, provided a
mathematical description of real world patterns, and a sufficiently
large amount of data \cite{Vespignani}. In fact, Takens' theorem
of chaos theory has been successfully applied to the prediction of the
behavior of real complex systems \cite{Sugihara}. Moreover,
typically there are advance warning signs of impending regime shifts, such
as a critical slowing down of relaxation times and an increase in autocorrelations
and variances \cite{EarlyWarningSignsNature}.

Socio-economic systems, however, are even more complicated to model.
Predicting the behavior of the economy or society
incorporates not only a higher degree of complexity, as the
aggregated behavior in such systems originates from the
dynamic interaction of a massive number of heterogeneous agents. The
agents also react to the knowledge of the forecast itself and adjust their behavior. 
In other words, once made public, a prediction becomes part of the
system dynamics. Nevertheless, a forecasting system considering the
response patterns of people seem to be possible, as lab experiments 
suggest \cite{Helbing_Lab_Experiments}.

Therefore, the huge amount of socio-economic data currently
accumulated from various sources \cite{Lazer} lets a number of leading
scientists believe that ``the complex systems that we are most likely
to tackle first in a truly quantitative fashion will not be the cell
or the Internet but rather society itself''
\cite{BarabasiDecadeScaleFree}.

\subsection{A distributed Center for risk analysis and crisis forecasting}

To cope with the new needs of societies to deal
with risk analysis and crisis forecasting, we envision the
establishment of an ICT Infrastructure in form of a distributed
Center on top of the existing European and National Grid
Infrastructures (EGI) along with the emerging Cloud Infrastructures.
Decentralized structures seem to be most adequate and to perform 
best in highly variable and heterogeneous complex systems such as network
traffic, ubiquitous computing systems (the Web of things), 
or smart grids \cite{Helbing_Lamer_selfcontrol}.

The overall scope of the multi-hub data Center should be to strengthen the
multi-disciplinary collaboration of the European research community of
complex systems scientists through the creation and deployment of services and
tools which will be built mainly upon the EGI Infrastructure 
with the aim of both facilitating new research groups joining the community and
increasing the research capacities and capabilities. 
Interdisciplinary science has received an increasing amount of attention
over the last two decades. Particularly due to changes in the rate
and nature of data collection in the physical, economic, social, climate,
seismological, physiological and biological sciences, new 
systematic approaches emerge. 

The distributed ICT infrastructure based on cloud or grid computing will permit associate
scientists to conduct data-analyses from a significant number of European research 
institutions, using the resources of the Center. This will be particularly
beneficial for scientists working at smaller universities, where researchers are
often lacking powerful computer equipment, particularly in the social sciences. 
Further proposals regarding novel ways to accelerate scientific progress
are made in the Visioneer white paper ``How to create an Innovation
Accelerator''\cite{VisioneerInnovationAccelerator}. The same white paper addresses, which ICT-based communication, coordination,
cocreation, quality assessment and management tools should be developed in order
to support an efficient collaboration of large numbers of scientists from different fields and institutions.

The Center should also issue special grants for young researchers from
different disciplines to encourage them to join the Center and work in
a goal-oriented manner. Calls for Ph.D. programs could be offered
world-wide and applicants would be evaluated by an
international scientific committee. Successful candidates would enter
a highly multi-disciplinary scientific career, which would guarantee
them extended visiting periods abroad at the other member
institutions. At the end of the program, they would be internationally
recognised as specialists in data-analysis.

Moreover, the tight collaboration between social scientists, ICT researchers and
engineers in the Center would be beneficial not only concerning
the analysis of the available data, but it would positively affect
also the methods and techniques of data-gathering. For example, new
embedded privacy-respecting solutions could be tried in a
well-defined test-bed and, if successful, successively implemented on
a larger scale. This would put social scientists into the position to
get exactly the data they need to advance their theories (as determined
in so-called ``Hilbert workshops'', which shall identify pressing research challenges). 
Furthermore, it is likely that the advanced research activities taking place in
this Center would give birth to new hardware or software
spin-off companies.

Specifically, the realization of such a Center for risk analysis and
crisis forecasting will

\begin{enumerate}

\item increase the usage of grid and cloud computing technologies by
  the training of potential users in targeted workshops, summer schools, etc., 
  \item develop Web-based scientific gateways providing intuitive access
  to the resources,
\item provide a common development framework providing users directly
  with frequently used algorithms and data curation tools,
\item build parallel versions (based on MPI) and hybrid versions (based on MPI and
  OpenMP) of basic algorithms, which will help to optimize the
  usage of the Center, 
\item foster theoretical progress of novel data-mining techniques
  through the development of new powerful computational algorithms
  able to automatically classify and distinguish spurious patterns
  from statistically significant ones,
\item build data repositories containing climate, social and 
  economic data relevant for risk analysis and crisis forecasting. A
  special data repository shall be built to collect and document stylized facts 
  of techno-socio-economic-environmental systems similar
  to the genetic databases available today,
\item promote a large-scale data-based approach in the socio-economic sciences,
\item attract brilliant minds to this new field of science by providing interesting good data
sources, new research challenges of practical relevance, and offering career perspectives 
in a multi-disciplinary and stimulating international environment, and
\item generate new hardware or software spin-off companies.

\end{enumerate}

The Center will pursue the integration of computational, storage and
data resources within the wider European Grid Infrastructure. It will
obviously requires a scale far beyond what has been used in the
socio-economic sciences so far. A collection of services and tools that
will help users benefit from this distributed infrastructure will be
designed and implemented. The sustainability of this Center will result from its 
decentralized and nationally rooted approach, but it will also depend upon the 
general usability of the resources and services and, thus, their thorough documentation,
interoperability, and integration---the implementation of the Center needs to
consider this.

\subsection{Towards predicting socio-economic crises} 

Many crises in techno-socio-economic-environmental systems are caused
by random coincidences or overcritical perturbations, which trigger
cascade failures (also known as domino or avalanche effects) in such a
way that the impact of random local events or perturbations becomes
systemic in size \cite{HelbingSystemicRisks_SE}. The sensitivity of
the system often results from the occurence of instabilities, which
create fertile ground for so-called regime shifts. Such regime shifts
happen at so-called critical points (``tipping
points''). Interestingly, when a system gets close to a tipping point,
it is often characterized by
\begin{enumerate}
\item slow relaxation (recovery) from perturbations,
\item increasing auto-correlations, and
\item critical fluctuations (a large variability). 
\end{enumerate}
Therefore, these features can serve as advance warning signs \cite{EarlyWarningSignsNature}.
Since they can be determined from empirical data, massive data mining will be able to
increase the level of awareness of upcoming crises and to trigger early preparations
in order to avoid or mitigate them. 
\par
While the above analyses are based on features of
phase transitions, methods from time series analysis may be applied as well. 
For example, Taken's theorem from chaos theory provides a method to identify
the mathematical regularities, which determine the dynamics in a complex system with
strong non-linear interactions (for which a classical time-independent statistical analysis mostly leads
to messy results) \cite{Sugihara}. Moreover, methods from statistics or statistical physics
can provide additional insights. For example, entropy methods allow one to measure
the degree of predictability of a system \cite{EbelingPocketsOfPredictability}. In addition,
analyses based on supply network models considering stock levels and flows facilitate the predication of 
shortages in resources \cite{Helbing_Lammer_supplynets}, which are often the basis of conflicts.
Finally, causality networks provide a method allowing one to anticipate possible courses of events, 
e.g. after a natural disaster strikes \cite{Helbing_Ammoser}.
\par
Techniques like these have not been extensively applied to anticipate and fight systemic crises in the 
past. They certainly promise better solutions for the future, supporting crises containment and the detection of feedback loops and possible cascading effects, before they cause wide-spread damage. They should be an integrative part of new ICT concepts for an adaptive risk management, facilitating and supporting a better disaster preparedness and response management.
\par
The following subsections mention a number of fields,
in which it would be possible to build up Crisis Observatories. While Crisis Observatories for different application areas could be created in different countries and institutions, it is obvious that problems in economic and environmental systems are interconnected. Similarly, the spreading of diseases does not only depend on virological and other biological factors, but also on the social behavior and the usage of the transport system, while it impacts the economy. Consequently, technical, social, economic and environmental systems are all mutually linked, forming one huge system that may be imagined as network of networks \cite{NaturePaperbyHavlinStanleyOnNetworksOfNetworks}. This makes it mandatory to foresee the future integration of the different Crisis Observatories, which requires the application of 
\begin{itemize}
\item the same standards, 
\item the creation of suitable interfaces, and 
\item interoperability. 
\end{itemize}

\subsection{Building up Crisis Observatories}

Many important changes in human systems as well as in most natural systems take place in rare crises or catastrophes that often appear as sudden and unanticipated ``storms''. The Crisis Observatories can be imagined as laboratories devoted to the gathering and processing of enormous volumes of data on both natural systems such as the Earth and its ecosystem, as well as on human techno-socio-economic systems, so as to gain early warnings of impending events. 

While national and European statistical offices (like EUROSTAT) are carefully studying {\it demographic changes} of the population structure (change of birth rate, migration, integration, effects of ageing on the health and retirement system, etc.), there seems to be potential for more intensive analyses in other areas such as financial and economic stability, social conflict, etc. The methodology of data-driven Crisis Observatories can, of course, also be extended beyond the range of examples given below, e.g. to observe cyber risks and critical infrastructures. 

\subsubsection{Financial and economic Crisis Observatory}

As the economic crisis has clearly revealed, it would be desirable to have a better picture of the stability of the fiancial system. Due to the strong systemic interactions, ``healthy'' looking balance sheets of most banks do not necessarily imply systemic stability. It is clearly necessary to have better models not only for each single aspect of the economy (such as the financial market, the housing market, the national economies, or the investment and consumption behavior on the microeconomic level). It is equally important to develop realistic systemic models, which integrate all of the above aspects and their non-linear interactions \cite{Econophysics,PluralisticModeling}. 
\par
While financial markets show a large degree of unpredictability, technical analyses are wide-spread and indicate that many players belief in their applicability. There seem to be a number of early indicators that may serve as warning signs, e.g. extremely high volatilities, increasing correlations in market activities, a quant meltdown (the sudden reduction in the leverage of hedge fonds), the liquidity of the interbank market, or the Arms index (a measure for the pressure to sell stocks). Further measures (for financial bubbles) have recently been proposed by Didier Sornette, who is establishing a Financial Crisis Observatory (FCO) at ETH Zurich \cite{FCO}. He has also detected bubbles in the housing markets in several countries in the past. On a much longer time scale, the gap opening between consumption and wages in the Unites States many year ago may have indicated the creation of market forces, which would sooner or later lead to a financial earthquake \cite{financialquake}.
\par
As a consequence, market monitoring seems to be promising to detect possible crashes in the financial and housing markets in the future. It would certainly make sense to analyze the financial flow network, considering besides the financial markets also factors such as debts and inflation, public spending, the sustainability of social benefit systems (and its dependence on the labor market), private investments (including the housing market), or impending shortages in resources (such as food, water, rare minerals [like Yttrium, Indium, Neodym, Tantal, Gallium, Kobalt], oil, etc.). Massive real-time data mining is expected to have an even greater impact, when combined with large-scale computational economic models.

\subsubsection{Crisis Observatory for conflicts}

A problem of similar severity as financial instability is the recurrence of conflicts in many areas of the world. While Europe has enjoyed a peaceful period for a long time, it has faced conflicts in former Yugoslavia and it has also been involved in international conflicts in Afghanistan and Irak. Europe is furthermore trying to solve the conflict in the Middle East, which has impacts all over the world. Besides, several countries in Europe are troubled by independence movements (France, Spain and Britain, for example), and some countries are experiencing social unrest in the wake of the current economic crisis (such as Greece and France). Last but not least, many European countries face problems with the increase of certain kinds of crime, with the social and economic inclusion of young people, and with immigration.
\par
This suggests to build up a crisis observatory, which maps, follows, and predicts conflicts in time and space. For this it should also evaluate factors provoking conflict, such as social, economic or political exclusion, the scarcity of resources, or the distribution of incompatible cultural values in space. In a multi-polar world, a particular focus needs to be put on the 
balance of power on different levels (countries, economic centers, cultures). Well-functioning democratic societies also need a suitable balance between individual and collective rights, central and decentral control, political and company power, etc. One furthermore needs to keep an eye on the situation of minorities, as their discrimination happens easily and often creates conflicts.

\subsubsection{Crisis Observatory for crime and corruption}

It seems also natural to establish a crisis observatory for crime and corruption. It would try to determine hot spots of crime and corruption, particularly their spatio-temporal dependencies. Furthermore, methods from network theory could be employed to reveal networks underlying organized crime and corruptions, and to help reduce their harmful effects on society. The method also seems to be promising to fight terrorism, drug dealing, and fraud networks, particularly as certain techniques allow one to reconstruct unknown parts of the networks.

Moreover, while collective social behavior can be powerful and positive, e.g. in creating public goods, our culture, and shared values, it can also assume pathological sides. Well-known examples for this are riots, hooliganisms, extremism, or terrorism. It may therefore make sense to track extremist opinions within public Internet forums, where attempts to mobilize support for illegal activities are made. This tracking will not be oriented at identifying individuals, but at finding hot spots triggering violent activity, which may be hard to detect without a systematic data mining. 

\subsubsection{Social Crisis Observatory}

The Social Crisis Observatory will be particularly focused on identifying factors that may sooner or later create dissatisfaction and conflict, such as social, economic or political exclusion of people of a certain gender, age, health, education, income, religion, culture, language, or preference. Similarly, it may be good to measure the level of social integration and factors influencing it. This information will allow one to determine the need for political action before social tension builds up to such a degree that it finally causes violent eruptions.

It could also be useful to measure factors sometimes summarized under the phrase ``social capital''. This includes factors such as solidarity, cooperation, trust, loyalty, compliance, which are relevant for the quality of life, but also for decisions regarding economic investments. Over-regulation, corruption, and other factors, however, which may speak against investments, would be equally important to measure. 

Further data-mining activities on an aggregate level may study factors such as consumption habits, travel behavior, recycling efforts, or energy use, which are relevant for an environmental-friendly behavior and its spreading. Going one step further, one may also consider to create something like ``moral sensors'' to identify changing norms. 
 
\subsubsection{Crisis Observatory for health risks and disease spreading}

Another significant application of Crisis Observatories are health risks. Not only diseases such as the flu, SARS, H1N1 or HIV are spreading by certain kinds of social interactions. Also obesity, smoking, or the tendency to commit suicide seem to be transmitted socially \cite{Christakis_obesity,Christakis_smoke,ChristakisReview}. It is obviously important to map these risk factors to health in space and time, to gain a better understanding of the mechanisms underlying their spatio-temporal spreading.  

Without any doubt, a particularly important, but also very challenging task is the real-time monitoring and prediction of the spreading of emergent diseases. Initially, the infection ways and rates of new diseases are not well-known. Hence, the parameters used in disease spreading models must be calibrated and adjusted over time. In the ideal case, this would happen automatically based on real-time infection data. Recent network-based measurement methods indicate the possibility of two-week forecasts \cite{Chr}, which promises major breakthroughs in terms of more efficient vaccination strategies.

However, the prediction of pandemic or epidemic spreading scenarios  over longer time periods will require to model also  human response to diseases. People change their behavior, which influences the spreading scenarios. Therefore, massive data mining will be needed to establish a behavioral epidemiology, facilitating more reliable forecasts than it was possible in the past.

\subsubsection{Transport and logistics Observatory}

For the first time in our history, the majority of humanity is living in an urban setting and this proportion will increase in the future. Congestion and property price bubbles are symptoms of coordination problems. Congestion generates losses of productive time, wastes energy, and pollutes the environment (particularly with CO$_2$). It creates economic losses of 10 Billion US\$ each year in the US alone and often makes transport and supply systems temporarily dysfunctional.

Congestion spreading has the typical features of failure cascading phenomena: Due to spill-over effects, certain links of the road network essentially fail to perform their transport task and affect further road sections. This can reduce the performance and throughput of an urban road network dramatically, while it may handle the same travel demand without any significant problems on other days. Traffic and supply systems are, therefore, an excellent example to study systemic failures based on cascade spreading. 

The Transport and Logistics Observatory is expected to support the smoother and environmental-friendly operation of traffic flows and supply systems. It could also reveal important properties of cascade spreading events, which can have system-wide impacts. Moreover, it should be able to provide powerful approaches for the short-term planning of large-scale evacuation scenarios, considering real-time feedbacks regarding actual threats and the traffic situation. Finally, the Transport Observatory should address the interconnections between traffic and land use, which are relevant for the development of real estate prices and industrial and urban development.

\subsubsection{Crisis Observatory for environmental changes}

Environmental changes have become a larger concern over the years, as their wider implications have been better understood. They trigger changes in soil, water, ecosystems, forests, and biodiversity. Moreover,
they have a relevant impact on food and water availability, flood and storm disasters, forest fires etc., which may affect huge numbers of people. Environmental changes affect the availability of energy, health, and wealth. They may cause agricultural degradation, economic decline, social unrest and political instability. 
This may impair the stability of societies, potentially causing riots, insurgencies, urban violence, or war.
\par
These circumstances require one to 
\begin{enumerate}
\item better understand the linkages between environmental change and
  social dynamics,
\item analyze the risks and conflicts that could affect societal and
  political stability,
\item assess how human beings and social systems respond to
  environmental change, 
\item detect and predict cascading effects and
  tipping points,
\item identify conditions for cooperation and sustainable problem
  solving.
\end{enumerate}
The above issues can only be addressed by mining Big Data. 

\section{Ethical and policy issues related with socio-economic data mining}\label{ethical}

Large-scale data mining is opening up previously unimaginable, new
perspectives for science (see Section \ref{sec_1}) and, of course, even more for business. 
At the same time, it affects fundamental rights of individuals in ways, which are 
hard to fully oversee. Among these, the right of privacy
is surely one of the most endangered, but it is not the only one. Such
risks result not only from single research or data-mining
activities. They arise in particular from the combination of singular
observations in larger datasets, which contain more and more
information, and are capable eventually to depict accurate personal
profiles. With these giant data conglomerates at one's disposal,
making sense of unpersonalized and apparently irrelevant information
is easier than one could think \cite{KleinbergAttacks}. However, it is
still not clear what the implications of developing such
\textit{informational cornucopias} are. In the meantime, the
construction, enlargement or acquisition of mega data centers run by
private companies and national security agencies spreads more and more
\cite{AppleDataCenter,NSADataCenter,MicrosoftDataCenter,GoogleDataCenter,GoogleDataCenter2}. Intel, the largest CPU
manufacturer in the world, has declared that already by 2012 mega data
centers will account for 20 to 25 per cent of its server chip sales
\cite{IntelFutureInDataCenters}.

In the following sections, we will discuss ethical aspects of building gigantic supercomputing ICT facilities for large-scale data mining, as the ones mentioned before. Our analysis will be primarily based on and guided by a literature review of ethical research in the social sciences. The approach followed can be characterised essentially (but not exclusively) by a positivist and structuralist standpoint, and our discussion will concentrate mainly on privacy issues. However, in section \ref{otherethicalissues} we will consider other ethical concerns inherently related to large-scale data-mining activities. Further ethical issues related to social super-computing are addressed in Ref. \cite{VisioneerSocialComputing}.

\subsection{A source-based taxonomy of available personal information}

Given that today, more information is available about us than we are
usually aware of, let us start the discussion of ethical issues with a
picture of the personal data traces almost everyone leaves most of the
time. The following paragraphs provide a non-exhaustive taxonomy of
available data organized by data-sources.

\subsubsection*{Data in public registries}

Data belonging this category is generally already public, or available
after the payment of small fees to public institutions.

\begin{itemize}
\item phone books,
\item land registries,
\item car plate registries,
\item health data,
\item salary registries (available primarily for the public
  sector)
\item tax data (public in the US),
\item religious confession,
\item social security and passport numbers.
\end{itemize}

\subsubsection*{Data generated by electronic services}

Today, the correct and efficient functioning of our everyday lives is
more or less dependent on a few essential services, which are
increasingly supported by ICT and electronic infrastructures. This
means that, by using such services, a lot of data are automatically
generated as by-product. Data in this category are usually
available only to certain public institutions and/or some private companies
providing these services.

\begin{itemize}
\item phone call logs,
\item flight passenger information (such as e-mail addresses, credit cards,
  etc., particularly for flights to the USA),
\item bank account data, 
\item credit card numbers,
\item money transactions (e.g. Swift system),
\item consumer data (``people who bought X have also bought Y''),
\item behavioural analyses.
\end{itemize}

\subsubsection*{Data generated by Internet activities}

``Look but do not touch'' was considered a wise advice to follow when
entering unknown environments. However, in the Internet, this is no longer
sufficient. The sheer surfing activity, without any content and without accessing any service requiring
authentication, e.g. reading a certain news, is enough to generate a wide range of
differentiated digital traces. These traces are stored on private
remote servers as well as on the local drives. This includes

\begin{itemize}
\item Internet service provider logs (e.g. IP and MAC addresses),
\item logs of remote access to phones and computers,
\item browser history,
\item browser cache,
\item cookies,
\item search queries, and
\item click streams.
\end{itemize}

\subsubsection*{Data from portable devices}

In many social strata, the everyday usage of portable devices is becoming a wide-spread
habit. The current integration trend makes
portable devices more and more interconnected with each other through
wireless communication networks. This facilitates the spatial tracking
of persons via location data, which are exchanged by their
devices. Such data include

\begin{itemize}
\item GSM, UMTS, and GPS location data,
\item WLAN/WiFi open hot-spots,
\item bluetooth devices,
\item RFID data,
\item car transponders for automated highway toll payment systems,
\item electronic badges (e.g. for conferences \cite{RFID}, hotel rooms, etc.)
\end{itemize}

Moreover, the large availability of peer-to-peer connections and Internet access points
increases the risk of security breaches and data leaks, especially when these devices are used by
people unaware of their vulnerabilities.

Finally, the portability of such media introduces the risk of loss of
the device itself and consequently of all data stored in them.
Given the ongoing miniaturization process and the steady
improvements in capacity, the privacy concerns arising from the lack
of encryption or other data-protection techniques for such devices are
real and concrete. This concerns, in particular,

\begin{itemize}
\item video and photo-cameras,
\item mobile phones,
\item electronic agendas and smart phones,
\item laptop,
\item flash memory cards and external hard drives, and
\item smart multimedia players.
\end{itemize}

\subsubsection*{Unauthorized content captured from diverse multimedia devices}

Individual actions that reveal the way of living of people may be
recorded in both public and private venues and made
public at any time and without any previous warning. This concern is
increasingly more concrete due to the integration of multimedia
contents into global projects such as Street View, and the
success of photo and video on-line repositories. This concerns

\begin{itemize}
\item uploaded content on social Web sites (e.g. embarrassing party
  snapshots or videos),
\item Google Street View photographs,
\item public webcams.
\end{itemize}

\subsubsection*{User-generated contents}

Many users ``voluntarily'' share personal opinions or even detailed
personal information on their on-line profiles. Whether they are aware
of all the risks of this practise it is not entirely clear, but the
material is sufficient to identify political, religious and/or sexual preferences of many Internet users.
This concerns

\begin{itemize}
\item blogosphere data (forums, blogs, chats, etc.),
\item the archive of mailing lists or discussion groups,
\item keyword scans of free mail accounts,
\item social network data. 
\end{itemize}

\subsubsection*{Security data}

Under the flag of security, people were willing or forced to reduce the
range of their personal freedoms, with consequences often also for
personal privacy. This can happen through an explicit disclosure of
personal data, e.g. filling in a security form to enter a foreign
country or through accessing a given service, or tacitly, e.g. through public 
surveillance cameras.

\begin{itemize}
\item video surveillance (CCTV),
\item face recognition data,
\item biometric data,
\item audio recordings, directional microphone observations, 
\item phone call surveillance,
\item speed radar photographs,
\item scanned items and body scans at airports,
\item security forms that must be filled in.
\end{itemize}

\subsubsection*{Intercepted data}

From very basic to very sophisticated techniques, despite this may be
for illicit purposes, electronic communications can be
intercepted. Examples include
\begin{itemize}
\item network eavesdropping (emails traffic, phone calls, etc.),
\item identity theft,
\item hardware trojans,
\item software trojans,
\item the physical analysis of variations in electromagnetic fields of
  wireless devices (keybord and mouse) and of computer screens,
\item the monitoring of fluctuations in the electricity consumption of electronic
  devices.
\end{itemize}

While the above lists are probably not complete, it is obvious that
the combination of only some of the above data can eliminate privacy
to a large extent. Modern information services give a striking picture
of this (see e.g. \cite{123people}).  On the one hand,
they show how much information can easily be gained about a single
person (contact data, pictures, videos, news, etc.). On the other
hand, they illustrate how easily wrong information not related to the
person searched for is mixed between correctly retrieved information.
Therefore, we will discuss below whether privacy is just an outdated
concept, or whether it is crucial for the functioning of democratic
societies.

\subsection{Why would the honest be interested to hide?}

When it comes to private data, some people suggest that privacy is
mainly in the interest of dishonest, criminal, or pervert people. In
the following, we will argue that this is a dangerous
misconception. Privacy has been granted not as a concession of the
state to the individual, but because a modern society needs it in order to florish.

Although different in several respects, commercial confidentiality may serve as useful illustration 
to explain why privacy is needed by individuals. For example, if
confidentiality would be dropped, there would be no incentive for
companies to invest into expensive long-term innovations, which pay off only 
through a winning margin. It would be so
much cheaper to copy inventions of others as soon as they occur.
(There would not be such a fierce discussion about copyright protection/patent enforcement, if
this would not be the case.) Secrecy and confidentiality are
needed to gain a competitive advantage (in particular in time) that makes innovation
commercially profitable. There
are two other interesting point about innovation:

\begin{itemize}
\item Innovation usually starts off in a minority position \cite{Mazloumiamn}. In the
  beginning, there are a few supporters and customers only. In other
  words, there is little innovation without the existence of minorities.
\item As is known from evolutionary theory, innovation thrives best
  when there is a large diversity of variants \cite{HelbingSaamTreiber}. In other words,
  diversity or ``pluralism'' is the motor driving innovation. Would we
  just orient ourselves at the majority or what is ``normal'' (the
  average), the innovation rate and, with this, adaptability to
  changing (environmental) conditions would be poor. This is actually
  the reason why totalitarian regimes are sooner or later destined to
  fail.
\end{itemize}

These principles can also be transferred to individuals. Without
privacy, pluralism is in danger, as the following lab experiment shows
\cite{LorenzRauhut}: Experimental subjects had to guess the correct
answer to a factual question such as ``How many murderers occurred in
the year 2006?'' They received a
certain amount of money, whenever their answer was close enough to the
correct one. In one setup, they decided several times without any
information feedback, in another setting, they were informed about the
estimates of the others. In the first round, the variation of answers
was high, but the correct answer was always within the range of
answers and was usually well approximated by the geometric mean value
of all estimates. When information feedback was provided, the answers converged over time, 
which may be taken as sign that the right answer was identified. Instead, however, it
often happened that the relevant spectrum of answers did not contain
the correct answer anymore. In other words, social imitation
created herding effects, which were often misleading.\footnote{Note that taking the wrong decision occurred here even without social pressure, while it is known since the famous Asch experiment that individuals give predominantly wrong answers (against their own judgement), if the people before them do so \cite{Asch}.} 

The financial crisis is probably an example for such herding effects, which led to
extremely expensive mistakes. Herding-related mistakes would become even more likely, when
people were put under pressure to conform with frequent opinions or
behaviors, and as the above experiment shows, even when they would only be
exposed more often to other opinions than they used to be. This applies as well to
many current Web services and recommender systems, which reinforce
dominating opinions.  
\par 
Revealing private data would increase this
tendency of conformism enormously and would have other unwanted side
effects, as the following points indicate:

\begin{enumerate}
\item If behaving ``normal'' would become the standard and
  individualism would be discouraged, life would become more
  predictable, but for sure also much more boring.

\item Conformism implies a danger of discrimination (for having a
  certain religious belief, age, gender, disease, sexual preference,
  etc.; it is not without reason that Americans apply for jobs without a
  birth date and without a photograph). It is well-known that minorities need protection. 
  One must be aware that it is usually minorities who create the concepts and
  life-styles of tomorrow, and that it is hard to say in advance,
  which ones will it be. The minority behavior that eventually wins a
  majority largely depends on environmental changes and historical
  developments.  A society, therefore, needs to
  have a pool of minorities to successfully adapt to the challenges
  and opportunities of the future. Minorities are an indispensable ingredient in the
  process for evolutionary innovation \cite{AxelrodEvoCoop,HelbingSaamTreiber}. 

\item The majority behavior of today may be a minority behavior of
  tomorrow. What is normal today may be perceived as abnormal
  tomorrow. For example, it is hard to predict how we will think 
  in the future about the appropriateness of certain kinds of food we eat or 
  the environmental and labor conditions under which goods that we buy are produced.
  Hence, nobody can be that sure his or her current behavior would 
  be considered proper in the future. Social norms are continuously changing
  \cite{NormsPaper}. For example, in the 60ies, the values of society
  were changing dramatically, and the establishment got under enormous
  pressure. There are many other examples, such as racial
  segregation, which was considered ``normal'' by many people in the past, but is seen in 
  a totally different way today. 
  
\item Private data could be misused by companies. For example,
  insurance companies have an interest to offer cheap contracts to the
  majority of people and to charge minorities for special risks
  (e.g. inborn or past health risks, or higher hospital costs of women
  giving birth to children). This, however, clearly undermines
  solidarity.

\item Publicly available data could be misused also for criminal
  purposes. For example, the city of Oakland releases information on
  where and when arrests are made, which is later on displayed on a
  private Web site \cite{OaklandCrime}. From that
  Web Site, it was possible for criminals to infer the police tactics,
  patrolling times and other valuable information
  \cite{EconomistDataDataEverywhere}.

\item Companies start charging money to people who want certain
  private information to be deleted
  \cite{Micropayments,InternetReputation1,InternetReputation2,InternetReputation3,InternetReputation4}.
  A recent newspaper article even predicts that privacy in future will
  be a privilege of the rich \cite{PrivacyForRich}.

\item Disclosing the wealth of people explicitly or implicitly (e.g. through Street
  View services) can endanger individuals and increases the chance that they
  may become victims of crime. Therefore, being rich may become less rewarding, 
  and all the private initiative, innovation and commitment leading to it as well. 

\item Generally, people with professions that require them to take
  unpopular decisions sometimes (such as judges, policemen, or
  teachers) need a certain degree of protection
  of their private sphere. Otherwise, they will not be able to exercise their
  job seriously anymore and end up doing what pleases those they have
  to judge.

\item People may lose the chance of forgiveness of the mistakes they
  have made, if information about them remains publicly
  accessible forever \cite{BestWayToClearYourName}. In the past, after a
  reasonable punishment, depending on the gravity of the
  misconduct, the policy of societies was to forget about them. In the worst
  case, delinquents could still emigrate to other countries, where nobody
  knew them, paying with the abandoning of their hometown a high price for
  getting a \textit{second chance}. Now, however, wherever one may go, the
  digital traces left behind will follow him or her. This is not
  necessarily bad, but it certainly requires a savvier society that is
  able to remember and forgive at the same time. As Thomas Szasz said
  ``The stupid neither forgive nor forget; the naive forgive and
  forget; the wise forgive but do not forget.'' Without an adequate
  mix of tolerance and solidarity, the ability of a society to
  (re-)integrate people could be seriously undermined. Outcasts would
  only have a chance to find friends among other outcasts. As a
  consequence, this would fragment society into a variety of subsocieties---a
  tendency, which can be observed already.

\item Whenever a huge amount of personal information is available,
  individuals, private businesses or public institutions may try to
  infer individuals behaviors, preferences and attitudes and to
  classify them according to certain profiles. This tendency is as
  strong as dangerous, since there is no such thing as an accurate
  classification. Moreover, in the presence of information
  asymmetries, which are extremely common in everyday life (such as
  market exchange, buyer/seller interactions, insurance contracts,
  bank operations, job interviews, etc.), an inappropriate or wrong
  classification may be hard to correct or oppose to. Moreover, it may
  affect the lives of people in manifold and unexpected ways, given
  the high degree of interconnectedness of different services. In the
  worst case, it can even drive people through no fault of one's own
  into {\it circuli vitiosi}, from which they cannot escape. For
  example, missing the repayment of the leasing of the car once could
  mark somebody as ``insolvent'' to the system. This label would
  prevent this person from getting future loans, which he or she would
  need in case of temporarily financial reverse. However, it could
  lead to even more paradoxical situations. For example, by skipping
  one installment, the system would automatically register the fact
  ``interruption of contract'', and tag one's profile with a negative
  label. Ironically, the real motivation behind the fact
  ``interruption of contract'' could even be that the whole amount
  of money due was paid at once, without waiting until the contract expired. 
  
  The above example is real, 
  and wrong classifications like these are
  already happening. But that is not yet the worst possible scenario
  imaginable. In fact, we must be aware that any form of
  classification introduces elements for discrimination, because the
  ``labels'' are often not fitting and not mutually agreed on
  \cite{WinterRauhutHelbing}. Classifications (be they justified or
  not) create peer groups and may seriously undermine the basis of
  cooperation and shared norms in our society. They may also cause
  unnecessary conflicts
  \cite{HelbingJohanssonNorms}.

\item As it becomes possible to learn quickly what kind of people we
  are interacting with and what they do and think, this will undermine
  an independent judgement of their qualities (and weaknesses). Rather
  than giving everybody a fair chance to find the right kind of
  friends, people might be put into a certain ``box'' and socially
  excluded. It is known that people need to be protected from intolerance, mobbing,
  blackmailing and bribery. To live in peace, people often choose to segregate
  themselves from others. Given the
  availability of a lot of personal information to everyone, however, the
  Internet does not allow this anymore. In this connection, it is important to note that
  undermining the mechanism of voluntary segregation can 
  seriously affect the cooperation among people, to the disadvantage of everybody \cite{HelbingYu,HelbingYuRauhut}.
 
\item The more the Internet knows about everyone, the closer we get to to a situation where
we can effectively read other people's minds. Such a situation, however, would
  potentially generate a lot more conflicts than we have today.

\item It must also be noted that having more information freely available does not
  necessarily lead to a more transparent, fairer or better society. In an information-rich environment,
  people spend only a short time on a certain subject, and it easily happens that people
  get a wrong impression and form unjustified prejudices based on such a {\it pars pro toto} approach (assuming that
  the first or a randomly picked piece of information would be representative for the full information). Therefore, 
  large amounts of information can promote misjudgements of somebody's behavior
  by the press and by the public opinion \cite{AgainstTransparency}. Such reputational effects are difficult to correct,
  particularly as rectifications (e.g. when a suspect in a crime case has been found innocent) are often poorly noticed.
  This may have a serious impacts on individual lives. 
  
\item When everybody has the same information at the same time (and
  at negligible costs), this may have negative feedback effects such as herding effects. A typical
  example is the information about a traffic jam, which is provided to
  everybody via the public news. One can easily imagine that this leads
  to over-reactions of drivers to the news and, thereby, to
  overcrowded alternative roads, while the originally congested route
  may become underutilized.  A possible solution of this problem is to
  provide user-specific information according to probabilistic rules
  \cite{Helbing_Lab_Experiments} or to overlay
  randomness to the information signal
  \cite{NewspaperArticleWithMarkusChristen2010}.

\item Systems where a high degree of transparency has already been
  implemented for years have shown to become more sensitive to sudden
  regime shifts. Examples are market hysteria and volatility
  clustering phenomena, which can cause failure avalanches. In some cases,   
  transparency on the producer side can also facilitate the
  establishment of tacit collusive practises, as it has been found in
  on-line markets, auctions, and laboratory experiments
  \cite{AuctionCollusion,ShultzTransparency,CollusionTransparency}.

\item Decisions to reveal private information may even spread in an ``epidemic'' way.
  For example, if someone decides to provide access personal data (such as GPS car 
  tracking data, in order to get a cheaper car 
  insurance contract), this can deteriorate the conditions and potentially narrow
  down the options for those, who do not want to give up their
  privacy.  In other words, revealing ones own data can have an impact
  on other people who did not like to do so, but who are eventually forced
  to provide private information in order to maintain the same contract conditions
  and the same price they had to pay before. This also applies to private health insurances, for example.

\item The data on the servers of certain Internet companies probably
  know more about us than our friends and partner, and maybe even
  better than ourselves. However, when knowing the preferences of
  customers, companies may try to manipulate their choices, and
  possibilities to do so may increase with personalized
  recommendations (special offers may even have addictive
  effects). As it becomes possible to shape the customers
  expectations, this is likely to decrease the willingness of
  producers to tailor products and services to the needs that
  customers really have. In fact,  due to the ``economies of scale'',
  businesses have a natural interest in providing a number of standard
  products. 

\item Going one step further, knowledge of the private preferences of individuals may be misused for targeted manipulation and exploitation,
which can finally result in a breakdown of trust into commercial offers and services. One must be aware that both, the financial and
commercial system of today are largely dependent on trust, and that damaging trust can seriously undermine the basis of future business
and its related profits.

\item Furthermore, recent scientific studies indicate that pluralism in a society may get lost, as new technologies change the parameters of the opinion formation dynamics \cite{HelbingFlacheMaes}. Socio-diversity and its benefits (as outlined above), may easily get lost in favor of conformism and monoculture. It requires the mechanism of individualization, i.e. the desire to be different from others. Therefore, technologies or circumstances promoting conformism may seriously endanger the basis of democracies. In fact, the danger to suppress minority opinions and preferences increases as large datasets containing private information are centrally stored, and as it becomes possible connect different kinds of datasets. It is clear that knowledge implies power, and it would be na\"{\i}ve to think that people would not use it. In fact, there are many examples of misuse of private data (see the section on cyber risks below). It would be surprising, if organized crime would not try to get access to Google's data. One of the few laws of social systems, which have been confirmed again and again is: ``Anything that can go wrong ... generally does go wrong sooner or later,'' This is concerning, as today's information systems  probably {\it would} give someone the power to damage today's pluralistic societies, if he or she really wanted. After all, the Internet contains more sensitive information and about a larger number people than secret services of totalitarian states ever had. 
In addition, experience tells us that no database is absolutely
safe. In 2009, for example, several large sensitive datasets have been
stolen from public institutions in Great Britain, where they should
have been well protected \cite{BritishIdentityStolen}.

\item Another important point to consider is that the public and private sphere are two sides of the same coin. Interestingly, public order cannot florish without the protection of privacy. Not only is privacy needed to provide personal freedoms as a compensation for the everyday submission to social norms in public life. Privacy is a valve for psychological pressure relief. Without private freedoms, everybody's thousandfold submission to a large number of rules and regulations of public life would be unbearable for many people, triggering rebellion sooner or later. The more rules people are asked to conform with, the more personal freedoms are needed to compensate for this. Finally, unveiling private life also means unveiling how many times people actually violate social norms. Recent scientific studies have shown that such knowledge undermines social norms and consequently leads to an erosion of social order \cite{RauhutDiekmann}. One prominent example is the sexual revolution in the 60ies, which was triggered by the publication of scientific studies revealing people's private sexual behavior \cite{Kinsey1,Kinsey2}.
\end{enumerate}

Therefore, the storage and processing of large datasets of
socio-economic activities is a very sensitive issue. They certainly have the
potential to harm pluralistic societies. The interests of individuals
(such as privacy) and companies (such as details of their business)
{\it must} be protected. Therefore, it is necessary to address cyber risks
and ethical issues by scientific, legal and technological means. The
following sections provide guidelines, how this could be done.

\subsection{Cyber risks and trust}

Big data aggregates represent much sought-after targets for cyber
criminals and big challenges for security experts. The Symantec
Internet Security Threat Report XV \cite{SymantecSecurityReportXV}
mentions a 100\% increase in the number of new malicious programs
identified (more than 240 million in 2009) and estimates the number of Internet users (companies and
individuals), who have been victims of cyber-attacks trying to steal
money or confidential information, to be of the order of 360 millions.
More and more attacks are aiming at \textit{identity theft}. Sixty
percent of all data breaches that revealed identities were in fact the
result of hacking.

An incomplete list of the risks one must be aware of when using the Internet today is given below:
\begin{itemize}
\item data theft,
\item theft of pin codes and passwords,
\item identity theft,
\item viruses, worms, and trojan horses (damaging software, steeling passwords, etc.),
\item data manipulation, 
\item wrong evidence (wrong accusations), 
\item damaging rumors \cite{RottenNeighbors}, 
\item information pollution, 
\item spam and unwanted advertisements.
\end{itemize}
These risks may seriously undermine the trust of people in the Internet and services based on it. For example, the theft of access data for electronic banking through phishing attacks has recently become a wide-spread problem. However, trust is essential for economic exchange. Systems which would effectively not work without a certain level of trust include:
\begin{itemize}
\item electronic banking,
\item e-mail,
\item eBusiness,
\item eGovernance, and
\item social networking.
\end{itemize}
To solve the above problems, the right mixture between legal regulations and technical innovations is needed.

\subsection{Current and future threats to privacy}

Whether personal data disclosure in the Internet is
the result of a truly \textit{voluntarily and deliberate} choice is rather
questionable. In social research, voluntarily participation is
considered a basic human right, which overlaps considerably with the
principle of informed consent \cite{Eu_Code_of_Ethics}. Moreover, European law, for example,
gives individuals an individual right of control over personal information. 
\par
There is no
unanimous definition for informed consent, but according to Diener and
Crandall\cite{Diener-Crandall} it is ``the procedure in which individuals choose
whether to participate in an investigation after being informed of the
facts that would be likely to influence their decision''. In
principle, any decision can be considered as informed consent if it
has been taken after being provided with the amount of information
that \textit{a reasonable and prudent person would want to
  know}\cite{FrankfortNachmias}. In the
Internet this is seldom the case. In fact, it is both possible and
relatively common for individuals to access Web sites without reading 
the terms and conditions (which may be several dozen pages long). It it also unlikely that most people would understand the full contract, while they actually have to approve this. 
Moreover, they are usually not given any options rather than accepting the conditions 
in order to get the requested service or rejecting them at the cost of no service, which does
not give users a reasonable choice. Under these circumstances, people may nominally 
give consent, but without being fully aware of or agreeing with the terms and conditions.
Such a situation would not be considered as informed consent \cite{WorkingParty}. This
contravenes a widely accepted principle in Social Science Ethics that
states that ``\textit{as far as possible, participation in sociological
research should be based on the freely given informed consent of
those studied}'' \cite{BSA,SRA}. Moreover, fully informing the
respondents it is not yet enough, since researchers should endeavour to
make sure that the participants of an experiment have fully understood
risks and consequences \cite{FrankfortNachmias}. This applies in particular for physically or mentally challenged individuals \cite{BSA,AoIR}, 
but cannot be ensured in the Internet \cite{WorkingParty}.

Whether large data-mining companies are aware of the above mentioned
ethical issues is questionable, especially when CEO's of big data mining companies
make statements about privacy such as the following
one: ``\textit{If you have something that you don't want anyone to
  know, maybe you shouldn't be doing it in the first place}''
\cite{GoogleCeoPrivacy}. This is worrying, because if ethical standards turned out to be 
insufficient at some of the fundamental places of command of the biggest data-mining
companies, or if market competition would push them to pursue only the
logic of profit, what would refrain them from collecting and using people's data
even in illicit ways? Data mining techniques improve every day,
while regulations and control over the gathered data are lacking far
behind. For example, tracking the source of collected
information---once it is stored in secured and not publicly accessible
databases---is virtually impossible; knowing who has access to which
kind of personal data is also not possible today. Relevant to this
discussion and particularly controversial is the latest case of
Street View cars. For several months, these cars have been
storing personal data, including passwords, credit card information
and accessed email contents, which were intercepted from private WiFi
networks. The incident was reported as result of a programming
error, but others have suggested that this was rather a case of WiFi sniffing,
as there exists a software patent which involves intercepting data and analyzing the
timing of transmission as part of the method for pinpointing user
locations. At the time of writing this White Paper, the actual
situation is still unclear \cite{GoogleCar1,GoogleCar2,GoogleCar3,GoogleCar4}.

Also when not possessing a sophisticated and expensive data mining system,
criminals can collect illicit data easily through Web browsers, as these
are daily affected by new malicious exploits (see \cite{SecurityFocus}). The most common attacks are now based on a technique
called ``history stealing''. Some Web sites even show this security
issue to visitors \cite{DidYouPorn,InternetKnowsYou}, thereby
demonstrating how easy it is to extract personal surfing habits of
Internet users. Scientific literature on the topic is vast, and latest
studies conducted on 243,068 users found that 76\% of them were
vulnerable to history detection by malicious Web Sites. Newer browsers
such as Safari and Chrome were even more affected, with 82\% and 94\%
of vulnerable users \cite{HistoryStealing}. Unfortunately, there is yet another privacy issue related to recent generations Web browsers: their inherently high customization capabilities have made them unique, and therefore trackable. In fact, even disabling cookies, and blocking history stealing-like exploits, individual Web surfing can still be reconstructed by simply following the customized ``fingerprint'', which the browser is carrying around from site to site. This fingerprint is actually made up by all the configuration information that the browser is exposing to remote Web sites. According to the Electronic Frontier Foundation \cite{HowUniqueBr}, information such as which plugins are installed, which fonts are available and which operating system the browser is running on, can create a unique portrait of 94\% of the visitors (for a self-demo see Ref. \cite{Panopticlick}).

Unethical or dishonest intents are not the only pitfalls glooming over
on-line data sharing. Even in a scenario, in which one has consciously
provided his or her own personal data to a company, which is using
them lawfully, unforeseen issues can suddenly arise. For example, such
a company could be sold or merged with another one, or simply, the
data could be sold, based on a change in the data handling
policies. Users are typically not notified of such changes, and they
usually have no effective possibility to draw back their data and
their consent to use them. Some social network Web pages are examples
for this.  In fact, because of the continuous updating and
modification of the terms of use \cite{Facebook_Privacy_Timeline}, the
Electronic Privacy Information Center (EPIC) has filed a formal
complaint at the US Federal Trade Commission \cite{Facebook_EPIC}, and
more lately US senator Charles Schumer (D-N.Y.) has petitioned the
Federal Trade Commission to request that the agency addresses the
issue of social network privacy policies
\cite{Facebook_US_Senate}. Moreover, some national data protection
commissioners have publicly warned of using certain social network
sites \cite{GermanWarnsFacebook,EuWarnsFacebook}.  Just recently, the
vulnerability of these services has been demonstrated by someone, who
downloaded 100 million user profiles and made them publicly available
for download \cite{100millionsFacebookUsers}.

Joining groups within social networks can offer another exploit for
potentially malicious de-anonymization attacks. A recent paper
\cite{DeanonimizeSocialNetworks} proved that 42\% of users that use
groups can be uniquely identified. This results are noteworthy,
because traditional privacy attacks were based on aggregating
information from multiple datasets. Such methods were based on
collaborative filtering \cite{CollaborativeFiltering} and enabled an
efficient and highly reliable characterization of a person from a few
data. The underlying technology is quickly advancing \cite{Candes},
and it may give service providers, such as mobile phone, Internet
television, or social gaming centers an unprecedented amount of
personal information. Research on related privacy issues and their
potential explicit or implicit consequences is still in its infancy
\cite{Privacy_Id_Management}. Moreover, an efficient legal protection
seems to be missing, while a simple-to-establish solution to some of
the above problems would be accountable pseudonyms \cite{Ziegler}.

Additional risks for the privacy of users emerge, when companies are
forced to reveal private data to governments or legal
institutions. Google offers a picture of the quantity of data which is
handed out to governments \cite{GoogleGovRequests}. There are also joints
startup companies with the CIA \cite{Google_Cia}. Finally, when data
are not subpoenaed or stolen from cyber-criminals outside of the
company, they can be leaked in the most fanciful ways, which go from
mislaying a physical device containing sensitive information, to the
dishonest action of a single employee from inside the company
\cite{AppleIpadLeak,BritishPhoneCustomerDataBreach}.

\subsection{Additional ethical concerns}\label{otherethicalissues}

Ethical problems are intrinsically ``ambiguous, uncertain and prone to
inevitable disagreement'' \cite{AoIR}, i.e the correct answer cannot be
deduced algorithmically from general rules to particular claims. They are related to
cultural values and social norms. In the following, we raise a number of open ethical questions connected with large-scale data-mining activities, to which, of course, we cannot provide definite answers here. Related research programs seem therefore in place. For the time being, governments and companies engaged in large-scale data-mining are advised to follow  the procedural ethics approach presented in section \ref{proceduralethics}.

\begin{itemize}

\item As large-scale data-mining activities are increasingly successful in predicting (aspects of) individual behavior \cite{SozialeNetzwerke}, 
they will constitute an extremely powerful tool. This raises issues of the possibility of misusing it. More importantly, it raises the question of who gets to use these tools on what grounds. Will it be national governments and international corporations? Would there be a moral imperative to make the systems available to developing countries, NGOs etc.? 

\item What about competing claims of systems? If an early warning system recommends certain activities, how should societies respond to such recommendations? For example, how to handle situations, in which a scarcity of resources occurs? 

\item Who will own the algorithms and the outcomes of the data mining activities? Intellectual property is often discussed in terms of ownership of data used for input, but the more interesting question would seem to be: who owns the predictions? As they could potentially be subject to patent protection for computer programs and business methods, a rigorous analysis of the implications of intellectual property protection for data mining activities is needed.

\item If policy is based on predictions, how open is the system to critical review? Who will know and understand the algorithms? How can mistakes in algorithms be identified and rectified?

\end{itemize}

\subsection{How to address ethical issues in large scale social data mining}\label{proceduralethics}

Large-scale data-mining raises both procedural and substantive ethical issues. Some of the latter are predictable and solvable by implementing legislative and technical solutions. In the case of privacy, for example, this would include

\begin{itemize}
\item the use of scanners for viruses, trojan horses, etc.,
\item encryption,
\item fragmentation of data \cite{WUALA},
\item restriction of access/read/write/execution rights (depending on the type of data and purpose),
\item selecting higher security standards in the browser (for example, turning off cookies or deleting the browser history),
\item anonymous surfing \cite{TOR,FREENET},
\item use of pseudonyms. 
\end{itemize}

Nonetheless, one needs to underline that a full understanding of substantive ethical issues would require a full knowledge of uses and applications of the system, which is impossible to acquire a priori. In order to ensure a future-oriented approach to ethics, every project performing large scale data mining should therefore incorporate procedures that will allow the identification of substantive ethics as well as ways of addressing them. Such procedures should include the governance of the project from inception to delivery and cover governance recommendations for the individual components (early warning systems). It should incorporate reflexivity in the project team, continuously discussing the following questions (and regularly seeking independent feedback from outside):

\begin{itemize}

\item What are the substantive ethical issues that can be foreseen at any given point in time? 

\item What are the assumptions underlying the project itself as well as those underlying the ethical analysis (what is perceived to be an ethical issue, and why?)

\item How can appropriate processes be established to address known ethical problems (e.g. informed consent procedures)?

\item How can factual knowledge about the product and its likely consequences be gained?

\item Who are the stakeholders affected by the system and how can their local knowledge be fed into the reflective process?

\end{itemize}

\section{Towards privacy-preserving data analyses}\label{privacy}

Privacy concerns, although often justified, can cause serious obstacles to socio-economic data mining, while in many cases such data-mining would be in the public interest, when done in a privacy-respecting way. For example, socio-economic data mining would be needed to gain a better understanding of socio-economic problems, how they arise and how they can be addressed. Therefore, the following sections elaborate concepts, how data mining could be done in a privacy-respecting way.

\subsection{Deliberate participation}

The simplest possibility to do social data mining is to do it with
data that individuals share deliberately. For example, some Web sites,
such as Blippy.com, Skimble.com or Swipely.com, collect everything
from consumer data over the last movie you have seen up how many push-ups you have done in your
last training session. Participants of these Web services intentionally make their data available to everybody, and they can be analyzed in any possible way. The only concern from a statistical point of view is that the set of people participating in these Web2.0 activities is not representative for the whole population, i.e. one would need to make complementary analyses in order to learn, how it is possible to correct for biases in these data. Typically, participants are younger than average and are not concerned to share their data because they lead pretty much ``average lives''. 
\par
Further data can, in principle, be analyzed by crawling the Web. The data out there are usually traces of, for example, shopping activities at eBusiness platforms or social networking activities. They are accessible to everybody in small numbers, and it is not clear whether and how much people would care about a company or person analyzing these data in large amounts, as they can be gathered by automated programs such as ``spiders''. There are certainly problematic applications of this kind, in particular when the resulting datasets are used to do business, although the data were not intended for this, or if they are sold to third parties with unknown intentions. 
\par
As the recent discussions about the activities of large data mining companies shows, legal regulations against unauthorised processing of individual data are urgently required. Scientific analyses, which lead to discoveries of public interest, may have a better justification, but it must nevertheless be decided in each single case, whether individual rights are touched and what is the public benefit of such analyses. Shear curiosity and the publication of a scientific paper may not be a sufficient justification, and therefore, the consultation of an ethical committee seems appropriate.
\par
As a consequence, it would be much better to work with data that people provide intentionally for a given purpose. Statistical samples can already be quite useful. Special ``on-demand-data-gathering'' tools could allow people to easily opt-in and opt-out of data-collection programs in a situation-specific way. For example, while people may usually object to provide their data, it is likely that the participation rate increases in special situations such as crises, where people tend to change their priorities and make a contribution. However, it is fundamental that the gathered data will be used only for the purpose people have explicitly given consent to. With the project ``Gaydar''\cite{Gaydar}, the MIT demonstrated how easy it is to filter out  from publicly available data sensitive personal information, which may be misused. This study predicted the sexual orientation of Facebook users by analyzing the publicly accessible pictures of their friends. Such studies suggest that the processing of data should be allowed only for a certain time period and for the purpose they have been provided, requiring that users have adhered to an explicit, fair, and informed opt-in procedure. For sensitive data-mining activities it would be appropriate to apply the standards followed in clinical studies today. In order to support on-demand participation, particular trust-worthy Internet platforms should enable the case-wise sharing of personal data according to the specified purposes. This could be a special function of future eGovernance platforms.

\subsection{Anonymization and randomization}\label{anonyrandom}

To satisfy the data protection directive 95/46/EC, any data containing personal information needs to be anonymized before it is evaluated. While this may be sufficient for many simple analyses, it may not guarantee that the identity of individuals cannot be revealed from anonymized datasets. Substantial research has been and is currently being performed in the database community on privacy preserving data mining, reflecting the importance of this subject \cite{DM_privacy,DM_privacy_2,DM_privacy_3,DM_privacy_4,DM_privacy_5} (for a comprehensive state-of-the-art summary see the  ``Privacy-Preserving Data Publishing'' survey \cite{PrivacyPre_Survey}). Nevertheless, there are still a number of open problems, and many approaches are standing next to each other, lacking user-friendliness, integration, and a consequent systemic approach. Problems occur in particular when datasets contain a list of many different features, and some combinations of features are rare. As a consequence, such data must be sufficiently coarse-grained and/or randomized to make sure that combinations of features occur in sufficiently large numbers and cannot be individually resolved. Furthermore, it must be avoided to save lists with many features in one single dataset. It is safer to store them separately on different computers and to access the separate datasets only with programs, which are guaranteed to determine coarse-grained properties only such as (sufficiently rough) statistical distributions. The resulting derivative datasets should be comparatively small and unspecific, or they should be surrogate datasets, in which the relevant {\it statistical} properties  are the same, but the underlying individuals (persons, companies, etc.) are randomly reshuffled and not identifiable anymore. 
\par
The generation of the anonymized, derivative, and surrogate datasets for the original data should be done by particularly qualified and trustable institutions, while a larger number of people can work with the resulting, less critical datasets. In the last decade, research in privacy-preserving data analyses has produced  methods and tools aimed at publishing data under a privacy-preserving shield. For example, data are made anonymous with respect to a certified trustable anonymity notion,€" which essentially guarantees that the probability of tracing back any data to the identity of the person to whom the data originally belongs is so low that it can be considered null in practice. Another active research line concerns the privacy issues in case of mobility data such as those produced by location aware devices \cite{DM_privacy_mobility,DM_privacy_mobility_2}. 

To protect the original datasets from theft and unauthorized access, the specially secured and authorized data centers should store them in an encrypted way, and decryption should be done only piecewise and for the miliseconds, when the derivative data are generated. All commands and source codes of computer programs involved in sensitive operations should be automatically protocoled on a separate server, which is unaccessible to persons who are authorized to deal with original datasets.

\subsection{Coarse-graining, hierarchical sampling, and recommender systems}

As indicated before, in case of sensitive data (such as pregnancy, religious confession, diseases or
the sexual orientation), it must be ensured that individuals and group
memberships cannot be identified from socio-economic datasets. For
this reason, datasets for statistical analyses must be coarse-grained
in a suitable way. This may also be done by real-time data-mining
(``reality mining'') approaches, if they are suitably designed. For
example, to determine congestion on a freeway, it is possible to
analyze mobile phone usage data, but it is not at all necessary to
know who is calling whom and what is the content. The same applies to
GPS localization information of mobile phones, if the distribution of
people is determined for the sake of an efficient evacuation. It is
just necessary to make sure that any potentially sensitive data (such
as the underlying phone number) is thrown away before the statistical
evaluation is performed. However, as the recent case of WiFi
recordings by Street View cars has shown, transparency is
needed for such applications, as one needs to make sure that really no
sensitive data are stored. In principle, it could be legally required
that the underlying algorithms are published, and no algorithms may be
used which are not open source.  
\par 
One particular approach in
reality-mining could be a hierarchical sampling via ad-hoc networks
of, for example, sensors or mobile phones, where detailed information
is only processed locally, and any transmitted information undergoes a
certain level of aggregation. That is, as data are distributed over
larger distances, they undergo several aggregation steps, which may be
imagined like a hierarchical sampling method. Whoever wants to process
a large dataset, would only get a coarse-grained view of the data,
since they would be accessible only via a high level in the
data-processing hierarchy. Whoever managed to see data on a lower and,
therefore, detailed level, would only have a very short-sighted and
limited view, i.e. see very little. It appears, however, that the
technical details of such systems matter in order to be sufficiently
privacy-protecting and acceptable to users of the resulting services
(e.g. location-based ones). A transparency of the data processing
algorithms and related legal regulations appear to be needed. It should be
explicitly forbidden and prevented to collect and store
low-aggregation-level data. It must be ensured that they are deleted
directly after they have been processed and before they are
transmitted. To be uncritical and widely acceptable, the processing
should happen in the technical devices used by the individuals and not
on company-owned infrastructures (as it is common today).
\par 
A possibility to make
low-level data robust to interception would look as follows: Given
that the data of interest can be represented as points in a
(quasi-)continuous space, one could add random numbers according to a
certain statistical distribution. Rather than transmitting the correct
value (such as the exact location of the individual), a random number
(``noise'') would be added, before the value is
transmitted to the ad-hoc network performing the reality mining. Such
random falsification would make low-level-aggregated data useless and
create a ``foggy'' situation that protects the individual from being
revealed \cite{Krum}.

However, if done in a suitable way, the aggregation of the individual data could
still lead to reasonably accurate results due to the law of large
numbers, according to which errors average out in a statistical sense.
\par Services of {\it recommender systems,} of course, need to target
an individual specifically, which seems incompatible with overlaying
noise. However, recommender systems could still be realized by applying a two-component
strategy: The first component would be a rough search, which does not
consider individual information or preferences (or only, when
sufficient noise is overlayed). Among the search results, the personal
computer or smart-phone of the user would then select the individually
fitting search hits, products, or advertisements, based on personal
information and preferences that are exclusively stored on the
individual computer rather than on a system of servers. Putting it
differently, recommender systems should be changed from an approach,
where individually customized recommendations are pushed to the 
user, to a pull approach, where
the user selects in confidence one option out of a larger spectrum 
of downloaded recommendations in
a way that does not reveal his or her preferences. Individuals who are
even concerned about storing personal information and preference data
on their own computational device should have the possibility to turn
off the second component, which would then result in untargeted
research results and in recommendations, which would not be individually
customized.
\par 
The same approach can be used in connection with location-based services, the great comfort 
of which many people do not want to miss anymore. 
Let us assume somebody wants to be guided to a erotic shop, but does not want the guiding
company to know this. The person would go to the center of town, and based on his or her
falsified, approximate position, the GPS location service would forward to the mobile phone
information about shops in the area. The user could then select among these according to
categories, but the selection would only be known to his or her phone. It would not be forwarded
to the content provider, nor would the exact location of the user be known to the provider.

\subsection{Multi-player on-line-games, pseudonyms, and virtual identities}

Another possibility to study social interactions are offered by
multi-player on-line games such as Second Life. The advantage of these
games is that players can participate under pseudonyms, without
revealing their real identity. From an experimental point of view,
this has some side effects, as people may behave differently under
anonymous conditions as compared to conditions with face-to-face
interactions. Still, these effects may be compensated for, and there
are a number of behaviors, which come out quite realistically. For
such reasons, studying interactions in multi-player on-line games is
becoming a research technique, which is used complementary to lab and
Web experiments \cite{FutureofSocialExperimenting,FutureofSocialExp_site}.  
\par 
Some of the artifacts of studying
multi-player on-line games result from the following facts (here we assume
that the system would not allow the registration of several {\it identical}
pseudonyms):
\begin{enumerate}
\item People may change identities, i.e. register as a new user if their previous behavior is sanctioned by other players or by the system (``whitewashing''). 
\item People may use multiple identities, potentially also in parallel.
\end{enumerate}
To overcome these problems, the following measures can be taken:
\begin{itemize}
\item Everybody could get a unique virtual identity, which would be needed to create unique pseudonyms. 
\item Registering a new identity could be made very time-consuming or costly.
\item People may be allowed to join a multi-player on-line community by invitation only (and there would be separate lists of members and pseudonyms, which would be secret and encrypted).
\end{itemize}
An additional problem is that people may buy an identity (pseudonym) with high reputation or scores from somebody else. This problem may be addressed by performing behavioral consistency checks to reveal the use of the same identity by different people. Alternatively or complementary, the matching of pseudonyms with the unique virtual identity could be sporadically checked (by requiring to enter it). 
\par
A unique virtual identity can be generated by a trustable public
institution such as the registration office. It is practically an
electronic signature that can be used to submit documents such as tax
declarations or payments. Note that there are already private
companies offering trusted virtual identities/electronic signatures,
among which Verisign, GeoTrust and Thawte.\par The unique
virtual identity would have a finite validity (i.e. it would have to
be regularly renewed), and plausibility checks for identity thefts
would be made, to invalidate stolen identities (such as for credit or
debit cards). The identities could, for example, be generated as
follows (where the system would log which administrative person handed out what card): 
When asking for the virtual identity, a box would be ticked
in the files of the respective person, indicating that a unique
identity card has been handed out. The identification number of this
card would be randomly generated, and the receipt of this card would be confirmed 
with a signature, showing a valid photo document (ID or passport). 
The identification number, however, would {\it not} be known to the office
handing out the card. 
\par
To reveal the real identity behind a virtual
identity in case of a severe crime, this should require the
simultaneous agreement of several independent authorities (e.g. judges
[who could also be from trusted non-profit organizations]). Only by
combining the keys of two thirds of the respective responsible
authorities, it would be possible to reveal the real identity. This
can be ensured by saving bits of the identity code in different
databases, all in an encrypted way. The access keys and responsible
persons controlling these keys would be regularly replaced by new
ones in order to avoid corruption. It would also be good to let 
computers randomly decide, which ones of a number of authorized persons
would have to decide whether to reveal the identity or not. This would
minimize external influence on the decisions of the respective
responsible authorities.

\subsection{Anonymous lab experiments}\label{Labex}

Social behavior can also be studied in lab experiments
\cite{Helbing_Lab_Experiments}. In these experiments, it may be desired to ensure
anonymity of the participants, as they may otherwise not reveal their
true opinions or their normal behaviors. In such experiments, it may
be needed that the experimental subjects do not meet the experimenter,
and maybe not even meet other experimental subjects.  
\par 
There are
different ways of implementing such a design. For example, individuals randomly
passing by an information stand could be invited to participate in the
experiment. If they were willing to participate, they would 
draw a lot with a unique number, and they would
enter the number of the lot into a time table, which is hidden from
the experimenter. At the time of the experiment, the experimental
subjects would show up in separate rooms, where they take their decisions
in isolation, based only on information coming from a computer
screen. Their decisions would then be transferred via Internet to the
other experimental subjects. After the experiment, the experimental
subjects would get an envelope with their compensation, which would be
pushed into their rooms through small slits under their locked
room doors. The subjects could leave their rooms two minutes later. The
experimental setup would ensure that nobody would know, who
participated in the experiments, and it would be unlikely that the
same person would participate several times. Nevertheless,
participants may be suspicious whether this setup will really be executed
in an anonymous way or whether there is a chance of hidden observation,
and this may still affect their behavior.  
\par 
A similar and even
more privacy-protecting setup can be realized via a Web experiment. A
large number of people would be informed that the experiment takes
place at a certain time and could log on with pseudonyms. Among the
people who have logged on the experimental Web page, the computer would
randomly match individuals to form experimental groups. At the end of
the experiment, each individual would get a voucher with a unique
code, which can be exchanged for the compensation for participating in
the experiment. One of the following ways of payment may be chosen
(listed in increasing order of anonymity):
\begin{enumerate}
\item The experimental subject gets the compensation from an
  independent cashier (e.g. the university cashier) against showing
  the voucher, without the need to sign a receipt.
\item The person gets the money from an independent, i.e. trusted third-party payment service (e.g. bank or post), when presenting the voucher (i.e. the voucher would basically be a cheque).
For example, it would be possible to use the mechanical turk \cite{MechTurk} 
for a third-party payment.
\item The experimental subject gets the compensation by entering the unique code of the voucher into a special cash machine.
\end{enumerate}
Experimental subjects could be recruited in different ways: The
simplest would be to display posters in public areas, calling for
participation at a specified time via a certain Web page (and people could actually participate from a computer in a computer pool or Internet cafe, if this gives them a better feeling of anonymity). Similarly, the announcement could be made via an advertisement on a heavily frequented Web portal. At the specified time, an algorithm would match visitors of the Web Site in a random way and try to make sure that groups of friends could not play with each other. 
\par
Avoiding that certain subjects participate in the same experiment
multiple times is more difficult (at least as long as only a few
people have unique virtual identities). One possibility would be to
send out invitations to a large number of e-mails, making sure that
there is only one e-mail address per person. People willing to
participate would enter into a Web page their e-mail address or a
unique code sent with the invitation e-mail. This is required for
authorization, to prevent multiple access. After this, they would be
redirected to a Web page, which shows a large list of unique codes, one
of which can be randomly chosen by clicking on it. This will cancel it
from the list and inform a Web service hosted by an independent, third
party (e.g. a computer center) that this code has been
authorized. When the participant enters the code into the Web page of the independent Web service, another code is returned, which is randomly selected from a long list of unique codes. That code will be needed to get access to the experimental platform at a later point. 
\par
The above procedure makes sure that the first step prevents multiple access. Afterwards, the selection of an individual code makes sure that the third party cannot have any clue of the relation between this chosen code and the e-mail address of the experimental subject. While it knows the list of acceptable codes, it does not know the identity of the person, just the fact that it is authorized to get a randomly chosen code from a list of unique codes, which are accepted by the experimenter. However, which code is randomly selected by the computer of the third party cannot be known by the experimenter. Finally, any temporal correlation among individual registrations is lost by implementing a sufficient time delay, after which the actual Web experiment takes place.

\section{Concept of a future, self-organizing and trusted Web}\label{newInternet}

In the following, we will describe technologies, which give people
back control over the data available about them. Some of the following
runs under the label of privacy enhancing technologies (PET). For
example, most Web browsers today allow one to turn off cookies (which,
however, makes certain Web services disfunctional). Furthermore, there
are tools such as Tor \cite{TOR} and Freenet \cite{FREENET}, which support
anonymous Web browsing and anonymized content sharing by obfuscating the IP address
of a computer. However, this is still not sufficient to guarantee
anonymous web browsing \cite{Panopticlick,BrowserFingerprints}.
Furthermore, one serious problem of today's
Internet still is the fact that it does not forget and that it does
not provide control over copies of data, which somebody has uploaded
in the past (e.g. party photos). First solutions for data with finite
lifetimes have become available only very recently
\cite{Vanish}. 

\subsection{Data format}

The following concept of a future, self-organizing and trusted Web is aimed
at overcoming the above mentioned and other problems. The basic
feature of the concept is a new ``Helbietti'' file format, which electronically signs and encrypts contents, but  has a number of unencrypted specifiers such as  
\begin{enumerate}
\item a unique file identifier (which is different for copies),
\item the kind of content (factual information, advertisement, opinion, unspecified), 
\item the lifetime (from ... to ...), 
\item a public annotation field allowing to tag the file and link it to others or to link it to comments or ratings, and
\item information regarding the price of producing and receiving a copy of the file.
\end{enumerate}
There would also be encrypted specifiers (readable only to authorized users), such as 
\begin{itemize}
\item the originator of the data (anonymous, pseudonym, real name, or company name),
\item the owner (anonymous, pseudonym, real name, or company name),
\item the date and time of generation,
\item a unique content identifier (e.g. check sum),
\item locations of authorized copies, and
\item the persons or groups authorized to read, modify, or execute the file (which would again be based on real names or pseudonyms etc., but one possible specification would be ``everybody'').
\item Annotations, which could be read only by the authorized persons or groups.
\end{itemize}
The following data would be double encrypted and accessible only to the owner of the file (and jointly to a specially authorized group of inspectors, see below):
\begin{itemize}
\item the file identifyer(s) of the file(s) it has been derived from (i.e. the previous version(s), if one existed, otherwise null) and the files that have been derived from it (e.g. any identical or modified copies),
\item all information regarding money transfers between customers or users of a file and the owner of its content as well as the respective tax authority, and    
\item the digital rights management settings (e.g. maximum number of
  copies that can be made from the original file).
\end{itemize}

To ensure that privacy and intellectual property rights are not undermined, checksum error-detection techniques would immediately reveal unauthorized manipulations to the original copy. Semi-automatic filtering measures could be implemented on servers which would refuse storage and forwarding of tampered copies. This kind of filtering may be compared to the immune defence system of the body against harmful viruses etc. For issues of copyright protection see Secs. \ref{copyright} and \ref{newInternet}. Moreover, depending on the sensitivity of the data (public, restricted, confidential, secret, etc.), they would be fragmented and distributed over several files stored in different locations \cite{WUALA}) and additionally password-protected, potentially requiring several passwords from independent authorized persons to access them. 

\subsubsection{Finite life-time data that can be controlled}\label{finicon}

This concept immediately allows one to limit the life time of data, as
they could only be decrypted within the specified time period. (Although 
there are first software solutions in this direction \cite{VanishWebSite,Vanish},
they seem to require further enhancements.) In order to avoid
tricking the file by modifying the time on a particular computer, the
file would automatically have to verify the time with one or several randomly chosen,
trusted servers (depending on the level of confidentiality; of course,
there would be a long list of such servers). Additionally, the file
could be opened in this time window only by individuals
or groups that are listed as authorized.

Besides, one could foresee a further restriction to the access of a
file by requiring that either the original file or one of the authorized 
copies are still accessible somewhere in the Internet. That
is, if the owner of the file would delete the original file and the 
authorized copies he or she may have created as
backups, no copies of the file may be opened any longer. This would
give the owner of the file perfect control over its distribution---a
fact which is also important for copyright protection (see Sec. \ref{copyright}).

\subsection{Intellectual property rights}\label{copyright}

The new data format also provides new possibilities to protect copyrights better. As music or video files would be encoded and require a certain password to open it, access to the file could be restricted to a single user or group of users. Moreover, Helbietti-formatted files could be set up in a way that a certain prize is charged (e.g. to a prepaid account) whenever a copy is produced. During the copy process, this amount of money would automatically be transferred to the owner of the intellectual property rights, the intermediate seller (e.g. a shop or the person whose file is copied) and the respective tax authorities. This would facilitate a ``viral marketing'', where users are distributors, who can earn money by disseminating file contents, while benefiting the owner of the intellectual property rights. This would, of course, not prevent the filming of videos and the illegal distribution of related copies. However, this problem could be minimized by a combination of the following measures:
\begin{itemize}
\item using pricing schemes that people consider fair,
\item selling copies of different quality at different prices,
\item allowing users to download contents with pseudonyms and
  anonymous payment services (e.g. \cite{Micropayments}), such that providers cannot track which contents are bought by what customers.
\end{itemize}  
Massive copyright violations could be reduced by using the labeling, reputation and sanctioning mechanisms described in Sec. \ref{newInternet}. Also note that the proposed file format allows one to make all copies unaccessible by deleting the single file that the copies were derived  from (see Sec. \ref{finicon}). Finally, for serious cases of piracy the new file format provides a possibility to track from what file a copy was derived, if decryption has been dcided by a number of specially authorized people (see Sec. \ref{trustmanagement}).

\subsection{Trust management}
\label{trustmanagement}
\subsubsection{Rating and reputation}

Many public goods such as reliable information systems 
are very difficult to create and easy to exploit and/or
destroy. This creates dangers for the quality of ``the commons'' (public goods).
In absence of clear responsibilities, such as it is often the
case in the Internet, large collaborative efforts are not encouraged. In
fact, contributors are more difficult to identify and to 
reward, while vandals and other detractors can easily thrive.

Therefore, the self-control of the Web,
based on suitable reputation concepts, would be a desirable feature.
In principle, people should be able to
rate, tag and comment on any data they have accessed. Also ratings and
comments could be rated, which would earn the rater a certain
reputation. Ratings would not be one-dimensional, but done on a
multi-dimensional scale (which could be customized in a user-specific
way). The multi-dimensionality is important to support pluralistic,
community-specific views.  

Note that details of the design of the rating mechanism are crucial. 
Manipulations of ratings must be prevented. The rating of
the raters can serve this purpose, if well constructed. It determines their weight in the
calculation of an aggregate rating. The design should be able to distinguish votes coming 
from robots and from humans. Furthermore, whitewashing (a
new pseudonym) and sybil attacks (the creation of many pseudonyms)
should be prevented (see the previous section regarding possible ways to do
this). Furthermore, to disclose a manipulation of the own reputation
via pseudonyms one is controlling (or a mutual manipulation through a
friendship network), consistency checks will be needed. That is, at
random times, it will be necessary to compare the reputation values
that a pseudonym has from the point of view of several others
(randomly chosen interaction partners, also independent
outsiders). This comparison of reputation values is something like a
gossip strategy. If the values are sufficiently consistent, everything
is fine and the reputation seems reliable. Otherwise, there are
reasons to be suspicious. In such a case, the pseudonym would be
labeled for the purpose of intensified observation in order to reveal
the manipulation. Such a differentiated inspection strategy, which
focuses on indviduals with a suspicious reputation (and newcomers),
but which otherwise restricts to random inspections, saves computational
resources but can keep the level of fraud low.  \par Furthermore, the
contents that users upload in the Internet would be rated by other
users who have access to them, earning the provider of the content a
certain reputation. This offers a tool to separate high-quality from
low-quality contents. In order to avoid opinion dictatorship by the
majority and ensure socio-diversity (pluralism), it will be necessary
to allow for community-specific and multi-criterial
ratings. Communities would either result from social networks, or they
could be determined via community detection algorithms, identifying
groups of people with similar rating, tagging, and commenting habits,
i.e. with similar preferences and tastes (so-called ``quality
collectives'' \cite{QLectives}). 
It should be remembered here that the
identities of the people belonging to a community will usually not be
known, but rather be composed of virtual identities, namely when
pseudonyms are used. 
\par 
The community-specific ratings, tagging and comments
can serve to create filters for certain contents. Therefore, it is
possible to design community-specific recommender systems which
prioritize contents fitting a community's or an individual's
taste. Similarly, undesired contents can be excluded so that it
becomes possible, for example, to protect children from sexually
explicit or violent contents. In other words, users could tag illegal
or inappropriate contents. In serious cases, this could trigger
sanctions (see below) or even legal action. For instance, the access
to the file could be restricted (e.g. to people above a certain age),
or the decryption could be disabled by a certain code foreseen by the
cryptographic algorithm. Also, in case there is evidence that access
to certain encrypted contents is in the public interest (e.g. relevant
for public security), the encryption method could foresee a
decryption. In order to avoid misuse such as censorship or violation
of privacy, both, the decision to restrict access and to enforce
decryption of a file or list of files, would need a certain number of
randomly selected, generally trusted and authorized people to agree on
the action that needs to be taken. Consequently, such actions would
require the application of several keys at the same time. To avoid
unjustified decryption by bribing authorized people, these should be
replaced after a certain time period, which means that the keys
unlocking a file need to change or be changed over time.

\subsubsection{Sanctioning mechanisms}

In reality, a reputation is hard to earn, but easy to lose. This suggests that, besides a reputation mechanism, the self-organizing Internet could foresee certain sanctioning mechanisms to facilitation a high level of quality. Sanctions may include everything from low ratings, over certain kinds of tags and critical comments, up to banning specific contents within a certain user community. Particularly destructive behavior may be sanctioned by temporary bandwidth reduction. For instance, manipulating ratings or reputation values by sybil attacks (self-ratings via multiple pseudonyms) should be sanctioned in one way or another. The same applies to wrong declarations (e.g. labeling advertisements or opinions as information). People should have a freedom to express their opinions, but they also need to have a chance to distinguish opinions from facts. Furthermore, spamming the Internet with low-quality contents should be sanctioned. Note, however, that what constitutes low-quality content for one community could constitute high-quality content for another community. That is, the sanctioning would usually be community-specific. Only in exceptional cases would it be generally applied.  

\subsection{Microcredits and micropayments}

It also seems wise to foresee in the future Internet the possibility to collect microcredits for small contributions to the public good ``Internet''. Such microcredits would allow one to reward people, for example, for contributions to public encyclopedias or also for rating contributions or reviewing (commenting on) them. 
\par
The data format of microcredits would, therefore, not only contain a certain value (``number of points''). It  would potentially also contain (usually in a sufficiently anonymized or encrypted way) information about who owns it and what it was earned for or paid for. Moreover, it would be a tradeable unit, which could contain pointers to who owned it last and whom it is being paid to (again in an encrypted way). Having both backward and forward pointers supports double book-keeping when needed. In mathematical terms, rather than a being scalar (which implies a number of fundamental problems), a microcredit would be an element of a microcredit network connecting values with pseudonyms and merits or items bought. These elements would have a certain number of weighted links (in-degrees and out-degrees) reflecting cash flows. Therefore, it would be possible, in principle, to distinguish different kinds of currencies for different kinds of contributions, and it would also be feasible to a certain extent to analyze flows of microcredits between pseudonyms over time in a privacy-respecting and confidentiality-protecting way (see the section on reality mining regarding how to do this; note that companies could use different pseudonyms for different organizational units, and that they may change them over time). Such kinds of analyses would be enormously useful to determine instabilities in the microcredit market. It would also be possible to give money a history and, therefore, distinguish ``dirty money'' (such as ``blood diamonds'') from ethical investments, as certain customers demand them today.

\subsection{Transparent Terms of Service}

In order to sign-up to a service in the Internet, one is more and more
often placed in front of a long list of obligations and contractual
clauses applicable for any sort of special case, which an ordinary
user has not the adequate competency nor the necessary time to
understand. The result is that they are skipped and blindly
accepted. Based on such ``acceptance'', companies can grant themselves
a great freedom of action in handling the personal data of their
users. This should not be allowed, and large-scale data mining
activities should be protocoled and publicly controlled.

Anybody willing to start collecting data from the Internet, or other
private and public nets should first publicly provide a legally binding
declaration about what is done with the data and why. In particular,
it should contain whom (what companies, institutions, etc.) the data
will be shared with, and what is done with them exactly.

This declaration should also comply with an international 
``\textit{data-collecting protocol}'', which needs to be established to set
legal and ethical constraints to the action of
\textit{data harvesters}.  The protocol should define minimum quality
of service standards, e.g. regarding waiting times of customer
services, times to delete private data, fees, how to contact the
data management center's service, whom to contact in case of
complaints.

Compliance to the protocol would allow companies to show a
``Privacy-Safe Badge'' on their Web site, which would immediately be
recognised by surfers (see \cite{PrivacySeals,IxQuick}). Showing the
privacy badge would probably become fundamental for certain categories
of companies operating the Net (e.g.  search engines, social networks,
banks, etc.). Not having the badge, could make a relevant difference
in the trust level of customers. Moreover, it is easily foreseeable,
as data-mining activities become more pervasive in the future, the
importance of the badge would eventually extend to other general
purpose Web sites.

The badge would be granted by newly created (ideally super-national) rating
institutions, which should also have the authority to enforce the standards related 
to the respective security badge by inspection. Such an institution
will be the only legal parties empowered to issue privacy badges and
revoke them in case of misconduct. In future, the collection of
personal data on the Internet without a proper badge could be
considered an illicit activity and insofar be sanctioned by users accordingly.

To obtain the badge, interested parties would have to demonstrate that they possess both, the
\textit{ethical} and \textit{technical} standards necessary to
accomplish such a delicate data mining task. After proper checking, and
depending on the purpose of the data collection specified in
the harvesting declaration, different types of badges could be
issued. Each badge would also be linked to a standardized user licence. 

In order to add dynamism and a more democratic taste, the badge could
foresee user ratings and comments. These opinion feedbacks per se, would 
not generate legal consequences for the owner
of the badge, but would help to detect misconducts earlier and to
alarm the community, and it would trigger inspection procedures by the issuer of
the badge.

Finally, users should be able to a-priori set their preferences and
conditions on their browsing devices under which participating or not
to data-collection campaigns\footnote{For example, willingness to
  allow collection of personal data only for scientific
  purposes.}. Browsers would immediately examine the badge of any
visited Web site, comparing it with its stored preferences and
automatically notify any threat to the privacy of the users. Besides,
this would solve the notorious issues of unreadable or over-technical
``Term of Services'' conditions, which should not any longer be read
directly by users themselves.

\subsection{Privacy-respecting social networks}

Social networking has been rapidly spreading in the past years despite
frequent warnings regarding a lack in privacy protection. Recently,
for example, somebody succeeded with downloading 100 Million user profiles
and upload the dataset for free download by everybody
\cite{100millionsFacebookUsers}. It is often claimed that users simply do
not seem to care about uploading private information to the Internet. However,
\begin{itemize}
\item this does by far not apply to everybody (in fact, most computer
  users still do not have social networking profiles),
\item the terms of use have changed since most of them have uploaded their private data (e.g. photographs),
\item some users may assess the comfort of the service provided by social networking sites higher than the current side effects, but this may change over time.
\end{itemize}
Besides, a recent empirical study has impressively demonstrated that people \textit{do}
care about the use of their activity data \cite{Palme}. 
\par
It certainly appears necessary to have alternative technical solutions
for social networking, which protect privacy better. A first project
of this kind is DIASPORA \cite{DIASPORA}, which wants to decentralize
the storage of sensitive information.  \par Privacy-protecting
social networks could be imagined as follows: Individuals would only
see part of the network. Individuals and communities could determine
what can be seen to outsiders of the community and to whom (friends,
friends of friends, second-next-nearest neighbors, or everyone; same
with business partners). Depending on this, certain kinds of
information would not be visible to outsiders, others would be (as
communities may want to gain new members). In essence, surfing in
social networks would be like travelling between communities, and this
would feel like visiting other countries. While certain things would
be visible, others (the private part of the information) would remain
hidden to strangers (as private houses are).

\subsection{Summary}

In essence, many of the problems of the Internet today result from Web2.0 and other applications, which the Internet was originally not designed for. Consequently, current technical solutions are insufficient. A new way of organizing the Internet appears to be needed and possible. Suitable solutions can be developed by transferring concepts of social self-organization to the design of the future Internet. This constitutes an interesting challenge within the research field of techno-social systems.

\section{Recommended legal regulations}\label{legal}

Currently, data about people are probably processed, used and misused in any conceivable way.
Since regulations are insufficient and heterogeneous,
the situation has sometimes been paraphrased as Wild Wild Web. 
It is therefore not surprising that the EU Fundamental Rights and Citizenship Commissioner Viviane Reding has
recently pointed out \cite{PrivacyHarmonization} that Europe needs
more harmonization regarding a data protection law. Determining the best routes towards this goal deserves
targeted research. However, as the problems are actute, action needs to be taken soon. Therefore, the 
following sections make a number of suggestions. 
\par
Given the problems of the current Internet and the foreseeable future
developments, data collection for research or for business should be regulated taking into account privacy, legal requirements, science's and business' interests. We foresee that methods of data collection should be open, controllable and verifiable by legal authorities and the public. Legal procedures and the law
should establish what type of data can be collected, what type of data may not be collected, and how the sensitive part of the collected data must be hidden
from people or organizations collecting them at each point of the data collection procedure. Methods of warranting the safety of sensitive data should be
public and should be verifiable at all times before, during, and after collection. For example, we recommend to work out legal regulations for the following:
\begin{itemize}
\item Data storage, access, processing and usage standards should be fixed for public, commercial and private entities operating in a certain country. Transparency regarding the storage, access, processing and use of data should be enforced. In particular, there should be a binding public declaration of what kind of private data are being stored, processed and used, and how this is done. 
\item Personal data should always be stored in an encrypted way. However, it should be made easy to inform oneself free of charge about the data determined and stored by other individuals, companies or institutions, and how these data are accessed, protected and used. Therefore, technical solutions should be required, which allow individuals to access (and decrypt) their personal data on-line and to delete data one does not want to be stored (if there is not a law that requires such storage). Further on, it should be easy to opt out from the determination, storage and/or processing of certain kinds of data. It should be possible to sanction violations of this right efficiently, and affected individuals should be properly compensated. 

\item One should establish standards ensuring informed consent of users with the data an information system is determining, storing, or processing. Users should not be forced to agree with a storage, processing and use of information that is not technically required for the services a user wants to get. For example, providers of media contents should not force customers to reveal their identities (and effectively their preferences via the contents they buy), as the contents or services can also be paid for anonymously. Putting it differently, companies should be required to offer, in a clearly marked way, options to customers that allow them to choose at any time between a data-rich variant (providing the service provider with many detailed individual data) and a data-poor (privacy-protecting) variant without artificially created disadvantages (which would effectively force customers to reveal their data). Within fair limits, it would be acceptable though to charge a higher price for data-rich services to users, who have decided for the data-poor variant themselves.

\item Licence and usage agreements of software products and information services should be regulated and standardized. As most users do not read or understand the terms of use, and as they do not have any chance to negotiate these conditions, there should be a few (certainly less than ten) standard licences, which should be signalized by a color or other codes, whenever a certain software or information-based service is used. Alternatively, softwares and services should be rated by independent agencies based on the benefits users can expect from them and the degree to which privacy and confidentiality are potentially affected.

\item It would be useful to define the individual and corporate responsibilities for damages created in the virtual or real world by activities in the Internet.
\item However, considering the fact that the content of a file is revealed only when it is accessed, it should not be possible to punish people for the access of contents, if the contents are not warned of in advance in a sufficient and qualified way (e.g. based on the rating and reputation system suggested above). In other words, users should be protected from legal traps in the Internet.
\item Considering the abundance of free contents in the Internet, it is advised to implement a copyright, which considers the facts of modern information systems and requires copyright holders to make proper attempts to protect their products from unauthorized access (e.g. to indicate their copyrights, encode electronic files, and offer simple, fair, and anonymous payment procedures).

\item Compensations for privacy violations would have to be fixed, and legal procedures would need to be simple and effective to allow people to protect their rights. For example, fines to companies, which sell private data without authorization, should be significantly higher than the likely profit they can make on such business.
\item The priorities in cases of conflicts of interest should be worked out clearly (protection of individual human rights comes before collective public interests, which comes before institutional interests of companies or political parties, which comes before individual interests).
\item Legal regulations should protect individuals against discrimination based on private data and guarantee an efficient compensation in case of violations. 

\item The introduction of class action would allow users to better defend their rights at court against individuals, companies, or institutions violating them, but the implementation should consider that the way attorneys are compensated and the way discovery is organized in civil procedures largely determines how desirable and effective class action lawsuits are. 

\item Unique virtual identities/electronic signatures should be offered for everybody.

\item It should be required to specially mark Web links that are redirected to contents with a different character, or Internet services that are changing their character (e.g. from non-commercial to commercial), or the use of pseudonyms that have been used before by others.

\item It may be useful to fix a statute of limitations, i.e. a time period after which violations of Internet-related regulations cannot anymore be sued. These time periods should increase with the seriousness of the violation and its consequences. It should also  depend on whether the effect of the violation was in the past or relevant for the presence and future as well.

\item There should be legal procedures regarding the random and targeted control of the fulfilment of legal standards regarding the storage, access, processing and use of private data.
\item Conditions should be worked out for imposing access restrictions or forced decryptions of suspicious Internet contents, in case there is evidence that they seriously threaten the public security (such as instructions how to build bombs). Such measures, their extent and results would have to be reported to the public, and individuals would have to be compensated, if it turned out that they were unjustified.

\item Companies receiving public money should be required to make data of public interest available for research, after they have been processed in a way that removes sensitive information (see the above sections on how this can be done).
\item It would be good to have neutral, publicly controlled third-party infrastructures, which allow to perform anonymous Web experiments and data mining.  
\item Special procedures should be defined for cases, where access to original or sensitive anonymized data is justified and required (e.g. for certain kinds of research of public interest). A good example is the way in which Harvard University regulates the access and processing of the data of the Framingham Heart Study, which allowed scientists to discover social processes promoting the spreading of obesity, smoking, depression, or happiness, to mention just a few relevant examples of gained insights that can be beneficial for individuals and the public \cite{ChristakisReview}.
\item There should be a fair right of information and participation in social activities mediated by ICT systems. For this reason, information businesses directed at a mass audience and with a large market share should be required not to discriminate and exclude certain user groups through inappropriate pricing schemes or terms of use (e.g. the requirement to agree with the arbitrary use or transfer of personal data or the requirement to allow for cookies, where this is technically not needed to provide the requested service). Individuals should always have the possibility to opt out of data uses they do not agree with, without losing access to information services not requiring these data.
\end{itemize}

\section{Recommended infrastructures and institutions}\label{infrastructure}

In order to have a powerful, largely self-regulating Internet, the following kinds of institutions would be useful to have:
\begin{enumerate}
\item Public data centers, which perform a neutral and independent data collection that is not driven by the need to make money, but serves the purpose to inform the public in the best possible way. Such a system could implement the reputation, community formation, sanctioning and privacy respecting mechanisms discussed before in connection with the concept of a self-organizing Internet.
\item Research centers, which study what can be done with publicly available data, to assess the potentials and risks. These centers should also develop the technology of the self-organizing Internet sketched above.
\item Publicly controlled, neutral institutions, which can serve as independent third parties in experimental designs that ensure anonymity (see Sec. \ref{Labex}).  
\item Independent quality audit centers, which evaluate the level to which companies protect privacy and provide good services and fair terms of use. 
\item One or several complaint center(s), which collects complaints of Internet users and can take action against illegal or unethical practices. These centers should be well connected with the public media.
\item An ethical committee, which assesses risks of information technologies and markets. It should set ethical standards regarding the storage and processing of data and support the preparation of required legal regulations.
\item A center working out contingency plans for the case of large-scale failures of information and communication infrastructures, e.g. due to denial of service attacks, spam, viruses, trojan horses, worm or phishing problems, or 
solar-storm-related failures of electronic systems. 
\item A committee working out suggestions for legal settings, as the need for institutional regulations arises through new technological developments.
\end{enumerate}

\section{Summary}

Socio-economic data mining has a great potential in terms of gaining a better understanding of problems that our economy and society are facing, such as financial instability, shortages of resources, or conflicts. Without large-scale data mining, progress in these areas seems hard or impossible. Therefore, a suitable, distributed data mining infrastructure and research center should be built in Europe. 

Reality mining provides the chance to adapt more quickly and more accurately to changing situations. For example, it will facilitate a real-time management of challenges like evacuation scenarios or economic stimulus programs. Further opportunities arise by individually customized services, which however should be provided in a privacy-respecting way. This requires the development of novel ICT (such as a self-organizing Internet), but most likely new legal regulations and suitable institutions as well. 

As long as such regulations are lacking on a world-wide scale (and potentially even thereafter), it is in the public interest that scientists explore what can be done (in a positive and negative sense) with the huge data available about virtually everybody and everything. Big data do have the potential to change or even threaten democratic societies. The same applies to sudden and large-scale failures of ICT systems. Therefore, dealing with data must be done with a large degree of responsibility and care. Self-interests of individuals, companies or institutions have limits, where the public interest is affected, and public interest is not a sufficient justification to violate human rights of individuals. Privacy is a high good, as confidentiality is, and damaging it would have serious side effects for society.

\subsection*{Acknowledgements}

The authors are grateful for financial support by the Future and Emerging Technologies programme FP7-COSI-ICT of the European Commission through the project Visioneer (grant no.: 248438).
They would also like to thank for stimulating discussions, feedback on the
manuscript and contributions to the Visioneer wiki: Karl Aberer, Andras
L\"orincz, Panos Argyrakis, Endre Bangerter, Andrea Bassi, Stefan Bechtold,
Bernd Carsten Stahl, Rui Carvalho, Markus Christen, Mario J. Gaspar da Silva, 
Fosca Giannotti, Aki-Hiro Sato, David-Olivier Jaquet-Chiffelle, Daniel
Roggen, Themis Palpanas, Elia Palme, J\"urgen Scheffran, David
Sumpter and Peter Wagner.

\appendix

\section{On-line repositories for the socio-economic sciences}
\label{Repository_Overview}

\subsection{Internet and historical snapshots}

\begin{itemize}

\item \textbf{Internet Archive / Wayback machine}

The Internet Archive offers permanent access for researchers,
historians, scholars, people with disabilities, and the general public
to historical collections that exist in digital format. Founded in
1996, now the Internet Archive includes texts, audio, moving images,
and software as well as archived Web pages.

\url{http://www.archive.org/index.php}

\item \textbf{Wikipedia} 

Wikipedia is the most famous cooperatively edited encyclopedia. Since
every change is stored, Web pages' history can offer a detailed
subject-based overview of the most important references of the past.

\url{http://www.wikipedia.org}

\item \textbf{The Knowledge Centers}

A collection of links to other resources for finding Web pages as they
used to exist in the past.

\url{http://www.searchengineshowdown.com/others/archive.shtml}

\item \textbf{Whenago}

Whenago provides quick access to historical information about what
happened in the past on a given day. 

\url{http://www.whenago.com}

\item \textbf{World Digital Library}

The World Digital Library (WDL) makes available on the Internet, free of charge and in multilingual format, significant primary materials from countries and cultures around the world. 

\url{http://www.wdl.org/}

\item \textbf{Books Ngram Viewer}

Timeseries for the two billion words and phrases, based on 5.2 million books written in seven languages.
\begin{itemize}
\item Datasets: \url{http://ngrams.googlelabs.com/datasets}
\item Application: \url{http://ngrams.googlelabs.com/}
\end{itemize}

\end{itemize}

\subsection{Information retrieval engines}

\begin{itemize}

\item \textbf{Freebase}

Freebase is an open, Creative Commons licensed repository of
structured data of more than 12 million entities. It provides
collaborative tools to link entities together and keep them updated.

\url{http://www.freebase.com}

\item \textbf{Wolfram Alpha Computational Knowledge Engine}

An attempt to compute whatever can be computed about
anything. It aims to provide a single source that can be relied on by
everyone for definitive answers to factual queries.

\url{http://www.wolframalpha.com/}

\end{itemize}

\subsection{Text mining on the Web}

\begin{itemize}
\item \textbf{Google Trends}

Google Trends shows visual statistics about how often keywords have
been searched on Google over time. Google Trends also shows how
frequently topics have appeared in Google News stories, and in which
geographic regions people have searched for them most.

\url{http://www.google.com/trends}

\item \textbf{Google Flu Trends }

Google Flu Trends uses aggregated Google search data to estimate flu
activity. Data available for download as well.

\url{http://www.google.org/flutrends/}

\item \textbf{The Observatorium}

The Observatorium project focuses on complex network dynamics in the
Internet, proposing to monitor its evolution in real-time, with the
general objective of better understanding the processes of knowledge
generation and opinion dynamics.

\url{http://www.theobservatorium.eu/}

\item \textbf{We Feel Fine}

A database of several million human feelings, harvested from blogs and
social pages in the Web. Using a series of playful interfaces, the
feelings can be searched and sorted across a number of demographic
slices. Web api available as well.

\url{http://www.wefeelfine.org/}

\item \textbf{CyberEmotions}

The CyberEmotions project focuses on the role of collective emotions
in creating, forming and breaking-up ecommunities. It makes available
for download three datasets containing news and comments from the BBC
News forum, Digg and MySpace, only for academic research and only after
the submission of an application form.

\url{http://www.cyberemotions.eu/data.html}

\end{itemize}

\subsection{Social data sharing}

\begin{itemize}

\item \textbf{Linked Data}

Linked Data is about using the Web to connect related data that was not previously linked, or using the Web to lower the barriers to linking data currently linked using other methods. 

\url{http://linkeddata.org}

\item \textbf{Dataverse Network Project}

The Dataverse Network is an application to publish, share, reference,
extract and analyze research data. It facilitates making data
available to others, and allows to replicate others work. Researchers
and data authors get credit, publishers and distributors get credit,
affiliated institutions get credit.

\url{http://thedata.org/}

\item \textbf{Data360}

Data360 is an open-source, collaborative and free Web site.  The site
hosts a common and shared database, which any person or organization,
committed to neutrality and non-partisanship (meaning ``let the data
speak''), can use for presentations and visualizations.

\url{http://www.data360.org/}

\item \textbf{Swivel}

Swivel is a Web site where people share reports of charts and numbers.
It is free for public data, and charges a monthly fee to people who
want to use it in private.

\url{http://www.swivel.com/}

\item \textbf{Many Eyes}

A IBM initiative that allows users to upload their datasets and use a
collection of tools to obtain meaningful visualizations from
them. Each visualization is publicly stored on a dedicated page, where
users can comment, rate and tag it. Reuse of the data is possible and
encouraged.

\url{http://manyeyes.alphaworks.ibm.com/manyeyes/}

\end{itemize}

\subsection{Conflict data}

\begin{itemize}

\item \textbf{CSCW Data on Armed Conflict}

CSCW and Uppsala Conflict Data Program (UCDP) at the Department of
Peace and Conflict Research, Uppsala University, have collaborated in
the production of a dataset of armed conflicts, both internal and
external, in the period 1946 to the present. Currently, probably the
most extensive dataset repository available, in particular for
historic data.

\url{http://www.prio.no/CSCW/Datasets/Armed-Conflict/}

\item \textbf{WarViews}

The aim of the WarViews project is to create an easy-to-use front-end
for the exploration of GIS data on conflict. It can run on a Web
browser or it can be displayed using Google Earth.

\url{http://www.icr.ethz.ch/research/warviews/}

The following are civil war specific datasets with additional
empirical information:

Ethnic group location dataset:
\url{http://www.icr.ethz.ch/research/greg}

Ethnic power balances dataset:
\url{http://www.icr.ethz.ch/research/geoepr}

\item \textbf{UCDP Datasets}

Collection of updated datasets and codebooks from the Uppsala Conflict
Data Program (UCDP).

\url{http://www.pcr.uu.se/research/UCDP/data_and_publications/datasets.htm}

\item \textbf{ACLED}

Partially contained in the PRIO dataset, ACLED (Armed Conflict
Location and Events Dataset) is designed for disaggregated conflict
analysis and crisis mapping. This dataset codes the location of all
reported conflict events in 50 countries in the developing world. Data
are currently being coded from 1997 to 2009 and the project continues
to backdate conflict information for African states to the year of
independence.

\url{http://www.acleddata.com/}

\item \textbf{CERAC}

The Conflict Analysis Resource Center hosts several cross country
conflict data sets and a few datasets of particular
countries. Repositories also have datasets of political instability
and conflict.

\url{http://www.cerac.org.co/datasets.htm}

\item \textbf{The Cross-National Time-Series Data Archive}

The Cross-National Time-Series Data Archive provides annual data for a
range of countries from 1815 to the present. Frequently cited, it is
one of the ``leading datasets on political violence'', according to Robert Bates at
Harvard University. It is ``possibly the most widely used event
dataset'' according to Henrik Urdal, International Peace Research Institute,
Oslo (PRIO).

\url{http://www.databanksinternational.com/}

\item \textbf{Country specific repositories}

\begin{itemize}
\item Iraq: \url{http://www.iraqbodycount.org and http://www.icasualties.org/}
\item Afghanistan: \url{http://www.icasualties.org/}
\end{itemize}

\item \textbf{Terrorism}

Collection of datasets of terrorist acts.

\url{http://people.haverford.edu/bmendels/terror_attacks}

\end{itemize}

\subsection{Data in economics and finance}

\begin{itemize}

\item \textbf{Bloomberg}

International real-time data provider for decision makers in finance,
business and government.

\url{http://www.bloomberg.com/}

\item \textbf{Maddison Data}

Historical statistics about GDP and population data.

\url{http://www.ggdc.net/maddison/}

\item \textbf{UNCTAD Statistics}

UNCTAD offers the following databases on-line:

\begin{itemize}

\item The UNCTAD Handbook of Statistics on-line provides time series
  of economic data and development indicators, in some cases going
  back as far as 1950.

\item The Commodity Price Statistics Online Database.

\item The UNCTAD-TRAINS on the Internet (Trade Analysis and
  Information System) for trade control measures as well as import
  flows by origin for over 130 countries.

\item The Foreign Direct Investment database (FDI).
\end{itemize}

\url{http://www.unctad.org/Templates/Page.asp?intItemID=2364&lang=1}

\item \textbf{OECD Statistics Portal}

Large collection of datasets covering economics,
demographics. Extractions are freely available, full access requires
subscription.

\url{http://www.oecd.org/statsportal/0,3352,en_2825_293564_1_1_1_1_1,00.html}

\item \textbf{EUROSTAT}

Detailed statistics on the EU and candidate countries, and various
statistical publications for sale.

\url{http://ec.europa.eu/eurostat/}

\item \textbf{Where's George?}

Spatial tracking system for U.S. and Canadian dollars.

\url{http://www.wheresgeorge.com}

\item \textbf{Eurobilltracker}

Spatial tracking system for Euro banknotes.

\url{http://en.eurobilltracker.com}

\item \textbf{EPO Worldwide Patent Statistical Database}

Snapshot of the EPO master documentation database (DOCDB) with worldwide coverage, containing 20 tables including bibliographic data, citations and family links. 

\url{http://www.epo.org/patents/patent-information/raw-data/test/product-14-24.html}

\item \textbf{World Bank}

The World Bank Data Catalog provides download access to over 2,000 socio-economic indicators from World Bank data sets. 

\url{http://data.worldbank.org/}

\item \textbf{Penn World Tables}

The Penn World Table provides purchasing power parity and national income accounts converted to international prices for 188 countries for some or all of the years 1950-2004.

\url{http://pwt.econ.upenn.edu/}

\item \textbf{World Input Output Database}

Exposes data about the effects of increasing globalization on trade patterns, environmental degradation and economic development that uncovers the global interrelatedness of production and its socio-economic and environmental effects. More in detail, data are available for the period from 1995 to 2006, and for some major countries back to 1980 27 EU countries and 13 other major countries in the world
More than 30 industries and at least 60 products

\url{http://www.wiod.org/database/index.htm}

\item \textbf{International Monetary Found}

The IMF publishes a range of time series data on IMF lending, exchange rates and other economic and financial indicators. 

\url{http://www.imf.org/external/data.htm}

\end{itemize}

\subsection{Scientific collaboration data}

\begin{itemize}

\item \textbf{ISI Web of Knowledge}

Comprehensive source of information in the sciences, social sciences,
arts, and humanities. It encompasses several datasets, among which
the following are maybe the most noteworthy:

\begin{itemize}
\item \textit{Journal Citation Reports}. It allows one to evaluate and compare
  journals using citation data drawn from over 7,500 scholarly and
  technical journals.
\item \textit{Web of Science}. It consists of seven databases containing
  information gathered from thousands of scholarly journals, books,
  book series, reports, conferences, and more.
\end{itemize}

\url{http://isiknowledge.com}

\item \textbf{Google Scholar} 

Google Scholar is search engine specialized in scholarly
literature. It indexes different sources (articles, books, abstract,
thesis, etc.) from several disciplines and sort them according to
number of citations, author and journal impact factor.

\url{scholar.google.com}

\item \textbf{Scholarometer} 

Scholarometer is a social tool to facilitate citation analysis and
help evaluate the impact of an author's publications. It works as a
software plug-in for the Firefox browser.

\url{http://scholarometer.indiana.edu}

\item \textbf{Scopus}

Scopus is a very large abstract and citation database of research
literature. It is available only for registered users.

\url{http://www.scopus.com}

\item \textbf{Living Science}

Living Science is a real time global science observatory based on
publications submitted to arXiv.org. It covers real time (daily)
submissions of publications in areas as diverse as Physics, Astronomy,
Computer Science, Mathematics and Quantitative Biology. Currently,
contents are dynamically updated each day.  Living Science is a
powerful analysis tool to identify the magnitude and impact of
scientific work worldwide.

\url{http://www.livingscience.ethz.ch/}

\item \textbf{PubMed}

PubMed comprises more than 20 million citations for biomedical literature from MEDLINE, life science journals, and online books. 

\url{http://www.ncbi.nlm.nih.gov/pubmed}

\end{itemize}

\subsection{Social sciences}

\begin{itemize}

\item \textbf{ICPSR of the University of Michigan}

ICPSR offers more than 500,000 digital files containing social science
research data. Disciplines represented include political science,
sociology, demography, economics, history, gerontology, criminal
justice, public health, foreign policy, terrorism, health and medical
care, early education, education, racial and ethnic minorities,
psychology, law, substance abuse and mental health, and more.

\url{http://www.icpsr.umich.edu/icpsrweb/ICPSR/}

\item \textbf{UK Data Center of the University of Essex}

The UK's largest collection of digital research data in the social
sciences and humanities.

\url{ http://www.data-archive.ac.uk/}

\item \textbf{Berkeley's UC DATA Archive}

UC DATA's data holdings are primarily in the areas of Political,
Social and Health Sciences.  

\url{http://ucdata.berkeley.edu/data_record.php?recid=6}

\item \textbf{The Economic and Social Data Service (ESDS)}

The Economic and Social Data Service (ESDS) is a national data service
providing access and support for an extensive range of key economic
and social data, both quantitative and qualitative, spanning many
disciplines and themes. It contains a map of additional datasets from
several European countries.

\begin{itemize}
\item \url{http://www.esds.ac.uk/}
\item \url{http://www.esds.ac.uk/findingData/map.asp}
\end{itemize}

\item \textbf{CESSDA}

Wide data collections including sociological surveys, election
studies, longitudinal studies, opinion polls, and census data. Among
the materials are international and European data such as the European
Social Survey, the Eurobarometers, and the International Social Survey
Programme.

\url{http://www.cessda.org/}

\item \textbf{Gapminder Data}

Gapminder is a popular technology and Web application for
cross-visualisation of trends in time series of data. It also opens an
archive of multiple datasets on diverse socio-economic indicators.

\url{http://www.gapminder.org/data/}

\item \textbf{World Value Survey}

The World Value Survey provides data about values and cultural changes
in societies all over the world.

\url{http://www.worldvaluessurvey.org/}

\end{itemize}

\subsection{Urban data}

\begin{itemize}

\item \textbf{Global Urban Observatory database}

The Global Urban Observatory (GUO) offers policy-oriented urban
indicators, statistics and other urban information.

\url{http://www.devinfo.info/urbaninfo/}

\item \textbf{Urban Observatory}

U.S. based datasets about wealth, innovation and crime across cities.

\url{http://santafe.edu/urban_observatory/}

\item \textbf{Urban Audit}

Urban Audit contains a collection of comparable statistics and indicators for European cities. Data for most recent years is missing at the time of writing.

\url{http://www.urbanaudit.org/}

\item \textbf{Globalization and World Cities Research Network}

The Globalization and World Cities Research Network (GaWC) promotes himself as the leading academic thinktank on cities in globalization. Several datasets are available for large cities networks.

\url{http://lboro.ac.uk/gawc/data.html}

\end{itemize}

\subsection{Traffic data}

\begin{itemize}

\item \textbf{NGSIM}

The Next Generation Simulation (NGSIM) program was initiated by the
United States Department of Transportation (US DOT). The program
developed a core of open behavioral algorithms in support of traffic
simulation, and collected high-quality primary traffic and trajectory
data intended to support the research and testing of the new
algorithms.

\url{http://ngsim-community.org/}

\item \textbf{Swiss Federal Roads Office FEDRO}

The Swiss Federal Roads Office offers a comprehensive overview on
traffic flows in Switzerland. Data are collected by permanent automatic
traffic counting stations and complemented by regular manual checking
since 1961.

\url{http://www.astra.admin.ch/verkehrsdaten/00297/index.html}

\item \textbf{TrafficData}

The aim of the International Traffic Database (ITDb) project is to
provide traffic data to various groups (researchers, practitioners,
public entities) in a format according to their particular needs,
ranging from raw measurement data to statistical analysis. ITDb
promotes a flexible traffic data provision format based on user needs
and standard habits. 

\url{http://www.trafficdata.info/}

\item \textbf{Clearing House for Transport Data}

The Clearing House for Transport Data in the German Aerospace Center
is the first point of contact for a quick overview of the available
data. It is targeted at both organizations who gather
transport-relevant data and those who wish to use the results of such
research. The information offered includes the preparation of detailed
metadata on the data sets, as well as notes on possible uses and
sources.

\url{http://www.dlr.de/cs/en/desktopdefault.aspx/tabid?669/1177\_read?2160/}

\item \textbf{Desweiteren das Regiolab Delft}

The regiolab-delft initiative started just after 2000 as a joint
project led by TU Delft in association with the Municipality of Delft,
the TRAIL research school, the Province of South Holland, the Ministry
of Transport and several industrial partners. The archived dataset
consists of over 6 years of 1 minute averaged speed and aggregate flow
data from densely spaced inductive loops on the freeway network in the
province of south Holland and other data from intersection
controllers, license plate detection camera's and much more.

\url{http://www.regiolab-delft.nl}

\item \textbf{RITA}
\label{rita}
The Research and Innovative Technology Administration (RITA) of the
U.S. Department of Transportation offers several datasets about
maritime, freights, airline, passengers, etc. traffic statistics.

\url{http://www.bts.gov/data_and_statistics/}

\item \textbf{ETH Travel Data Archive (ETHTDA)}

The ETH Travel Data Archive (ETHTDA) is a virtual platform allowing
end users to browse the archived travel data over the Web and enabling
simple statistical analysis. 

\url{http://www.ivt.ethz.ch/vpl/publications/ethtda}

\item \textbf{Metropolitan Travel Survey Archive}

The Metropolitan Travel Survey Archive to store, preserve, and make
publicly available, via the Internet, travel surveys conducted by
metropolitan areas, states and localities.

\url{http://www.surveyarchive.org/}

\item \textbf{Infoblu}

Infoblu is a private company providing real-time traffic monitoring
services for Italy. All services are available for a fee. 

\url{http://www.infoblu.it}

\item \textbf{ENAC}

Our Air Transport database comprises rich and detailed information on
airlines, airports and traffic flow. In order to increase its scope
and its reliability, ENAC also carries out annual surveys of airlines
and airports.

\url{http://www.enac.fr/recherche/leea/databaseA.htm}

\end{itemize}

\subsection{Open maps}

\begin{itemize}

\item \textbf{Google Maps}

World-famous map service. It offers several additional services such
as: Street View, user-uploaded content (photos, comments and ratings)
and personalized overlays through service apis.

\url{http://maps.google.com}

\item \textbf{OpenStreetMap}

OpenStreetMap (by UCL) is a free editable map of the whole
world. OpenStreetMap allows you to view, edit and use geographical
data in a collaborative way from anywhere on Earth.

\url{http://www.openstreetmap.org/}

\item \textbf{Tracksource Brasil}

Tracksource is a collaborative project aimed at creating and
distributing for free maps of Brasil.

\url{http://www.tracksource.org.br}

\end{itemize}

\subsection{Logistics data}

\begin{itemize}

\item \textbf{National Household Travel Survey}

The National Household Travel Survey (NHTS) collect data on both
long-distance and local travel by the American public. The joint
survey gathers trip-related data such as mode of transportation,
duration, distance and purpose of trip. It also gathers demographic,
geographic, and economic data for analysis purposes. It is part of
RITA (\ref{rita}).

\url{http://www.bts.gov/programs/national_household_travel_survey/}

\item \textbf{Commodity Flow Survey}

The Commodity Flow Survey (CFS) is the primary source of national and
state-level data on domestic freight shipments by American
establishments in mining, manufacturing, wholesale, auxiliaries, and
selected retail industries. Data are provided on the types, origins
and destinations, values, weights, modes of transport, distance
shipped, and ton-miles of commodities shipped. It is part of RITA
(\ref{rita}) and it is conducted every five years (last sampling on
2007).

\url{http://www.bts.gov/publications/commodity_flow_survey/}

\end{itemize}

\subsection{Health Data}

\begin{itemize}

\item \textbf{World Health Organization}

The World Health Organization publishes on line several statistics and supply direct access to four rich databases:
 
- Global Health Observatory
- WHO Global InfoBase
- Global Health Atlas
- Regional statistics
 
\url{http://www.who.int/research/en/index.html}

\end{itemize}

\subsection{Climate and Environmental data}

\begin{itemize}

\item \textbf{J\"ulich }

Climate data from J\"ulich Research Center. \textbf{I did not find
  data available for download!}

\url{http://www.fz-juelich.de}

\item \textbf{Google.org}

Google introduces its data-driven philanthropic projects, among
which two environmental satellite observatories:\\
- the Earth Engine: for monitoring trends in world deforestation;\\
- the Crisis Response: for monitoring the oil spill from the Deep Horizon
sank platform.

\url{http://www.google.org/}

\item \textbf{Footprint Network}

Ecological Footprint and the biocapacity results for more than 100 nations, based upon data from 2007, the most recent year for which source data are available. The tables reflect the calculations from the 2010 National Footprint Accounts.

\url{http://www.footprintnetwork.org}

\item \textbf{PSD Climate and Weather Data}

PSD archives a wide range of data ranging from gridded climate datasets extending hundreds of years to real-time wind profiler data at a single location. The data or products derived from this data, organized by type, are available to scientists and the general public at the links below.

\url{http://esrl.noaa.gov/psd/data/}

\item \textbf{EPA DataFinder}

The Environmental Protection Agency Data Finder is a single place to find EPA's numerical data sources so that people can access and understand environmental information. All of the data sources are available on the Internet and have been organized by topics such as air, water, and chemicals.

\url{http://www.epa.gov/datafinder/}

\end{itemize}

\subsection{Energy}

\begin{itemize}

\item \textbf{International Energy Agency}

Vast repository of statistics about supply and consumption of energy sources. Some datasets available only for sale.

\url{http://iea.org/stats/index.asp}

\end{itemize}

\subsection{Governance, Trade and Settlements Data}

\begin{itemize}

\item \textbf{Govindicators}

The Worldwide Governance Indicators (WGI) project reports aggregate and individual governance indicators for 213 economies over the period 1996–2009, for six dimensions of governance:
\begin{enumerate}
\item Voice and Accountability
\item Political Stability and Absence of Violence
\item Government Effectiveness
\item Regulatory Quality
\item Rule of Law
\item Control of Corruption
\end{enumerate}

\url{http://info.worldbank.org/governance/wgi/index.asp}

\item \textbf{WTO International trade and tariff data}

The World Trade Organization offers an updated and comprehensive outlook over trade policy and multilateral trading systems.

\url{http://www.wto.org/english/res_e/statis_e/statis_e.htm}

\end{itemize}

\subsection{Reality mining}

\begin{itemize}

\item \textbf{Reality Mining}

Behavioral data collected from 100 mobile phones over 9
months. Includes both proximity and phone usage statistics. Two
anonymized datasets available: single user (MySQL) and global
(Matlab).

\url{http://reality.media.mit.edu/}

\end{itemize}

\subsection{Other open data initiatives}

\begin{itemize}

\item \textbf{Data.gov}

Wide collection of public US datasets available for research.

\url{http://www.data.gov}

\item \textbf{Data.gov.uk}

Wide collection of public UK datasets available for research.

\url{http://data.gov.uk/}

\item \textbf{Digging Into Data}

Launched by the National Science Foundation (NSF), it offers a
collection of diverse data repositories.

\url{http://www.diggingintodata.org/}

\item \textbf{Guardian Data Blog}

Data journalism initiative that posts public interest (primarily UK
relevant) datasets together with their analysis. A few collaborations
with data visualization artists are present as well.

\url{http://www.guardian.co.uk/news/datablog}

\item \textbf{Google Public Data}

Google offers several large datasets on diverse world socio-economic
indicators and provides tools for easy visualization.

\url{http://www.google.com/publicdata/}

\end{itemize}

\end{document}